\documentclass[conference]{IEEEtran}
\IEEEoverridecommandlockouts
\usepackage{cite}
\usepackage{amsmath,amssymb,amsfonts}
\usepackage{algorithmic}
\usepackage{graphicx}
\usepackage{textcomp}
\usepackage{xcolor}

\usepackage{balance}  
\usepackage{comment}
\usepackage{cleveref}
\usepackage{bm}
\usepackage{makecell}
\usepackage{stackengine}

\usepackage[font=small]{caption}

\usepackage{latexsym}
\usepackage{bbding}
\usepackage{array}
\usepackage[ruled,vlined,linesnumbered]{algorithm2e}
\usepackage{subfigure}
\usepackage{multirow}
\usepackage{makecell}
\usepackage{enumitem}
\usepackage{diagbox}
\usepackage{setspace}
\usepackage{url}

\SetKwRepeat{Do}{do}{while}
\SetKw{Continue}{continue}

\newtheorem{myExample}{Example}
\newtheorem{myDef}{Definition}
\newtheorem{myTheorem}{Theorem}
\newtheorem{myLemma}{Lemma}
\newcommand{\tabincell}[2]{\begin{tabular}{@{}#1@{}}#2\end{tabular}}
\makeatletter
  \newcommand\figcaption{\def\@captype{figure}\caption}
  \newcommand\tabcaption{\def\@captype{table}\caption}
\makeatother

\def\BibTeX{{\rm B\kern-.05em{\sc i\kern-.025em b}\kern-.08em
    T\kern-.1667em\lower.7ex\hbox{E}\kern-.125emX}}

\begin{document}

\title{Aggregate Queries on Knowledge Graphs: \\ Fast Approximation with Semantic-aware Sampling}

\author{
\IEEEauthorblockN{Yuxiang Wang\textsuperscript{1}, Arijit Khan\textsuperscript{2}, Xiaoliang Xu\textsuperscript{1},
Jiahui Jin\textsuperscript{3}, Qifan Hong\textsuperscript{1}, Tao Fu\textsuperscript{1}}
\IEEEauthorblockA{\textsuperscript{1} \textit{Hangzhou Dianzi University, China}
\textsuperscript{2} \textit{Nanyang Technological University, Singapore}
\textsuperscript{3} \textit{Southeast University, China}}
\{lsswyx,xxl,qfhong,taof\}@hdu.edu.cn, arijit.khan@ntu.edu.sg, jjin@seu.edu.cn
}

\maketitle

\renewcommand{\IEEEQED}{\IEEEQEDopen}
\def\IEEEproofindentspace{2\parindent}
\renewcommand{\IEEEproofindentspace}{10pt}

\begin{abstract}
A knowledge graph (KG) manages large-scale and real-world facts as a big graph in a schema-flexible manner. Aggregate query is a fundamental query over KGs, e.g., ``{\em what is the average price of cars produced in Germany?}''. Despite its importance, answering aggregate queries on KGs has received little attention in the literature. Aggregate queries can be supported based on factoid queries, e.g., ``{\em find all cars produced in Germany}'', by applying an additional aggregate operation on factoid queries' answers. However, this straightforward method is challenging because both the accuracy and efficiency of factoid query processing will seriously impact the performance of aggregate queries. In this paper, we propose a ``sampling-estimation'' model to answer aggregate queries over KGs, which is the first work to provide an approximate aggregate result with an effective accuracy guarantee, and without relying on factoid queries. Specifically, we first present a semantic-aware sampling to collect a high-quality random sample through a random walk based on knowledge graph embedding. Then, we propose unbiased estimators for COUNT, SUM, and a consistent estimator for  AVG to compute the approximate aggregate results based on the random sample, with an accuracy guarantee in the form of confidence interval. We extend our approach to support iterative improvement of accuracy, and more complex queries with filter, GROUP-BY, and different graph shapes, e.g., chain, cycle, star, flower. Extensive experiments over real-world KGs demonstrate the effectiveness and efficiency of our approach.
\end{abstract}

\section{Introduction}
\label{intro}
Knowledge graphs (KGs) are popular in managing large-scale and real-world facts \cite{Huang2019,Gao2019}, such as DBpedia \cite{Lehmann2015}, YAGO \cite{Hoffart2013}, Freebase \cite{Bollacker2008}, and NELL \cite{Mitchell2018}, where a node represents an entity with attributes, and an edge denotes a relationship between two entities. Querying KGs is critical for a wide range of applications, e.g., question answering and semantic search \cite{Guha2003}. However, it is challenging due to the KG's ``\textit{schema-flexible}'' nature \cite{Yang2014,Zheng2019,Zheng2017,Janev2020,Zheng2016, Wang2020}: \textit{The same kind of information can be represented as diverse substructures} \cite{Zheng2016, Wang2020}.
This schema-flexible nature should be carefully considered in the study of KG querying, especially for the following two important query forms: \textit{factoid query} and \textit{aggregate query} \cite{Cui2017}.

\begin{figure*}
\vspace{-0.4cm}
\setlength{\abovecaptionskip}{0.1cm}
\centerline{\includegraphics[scale=0.66]{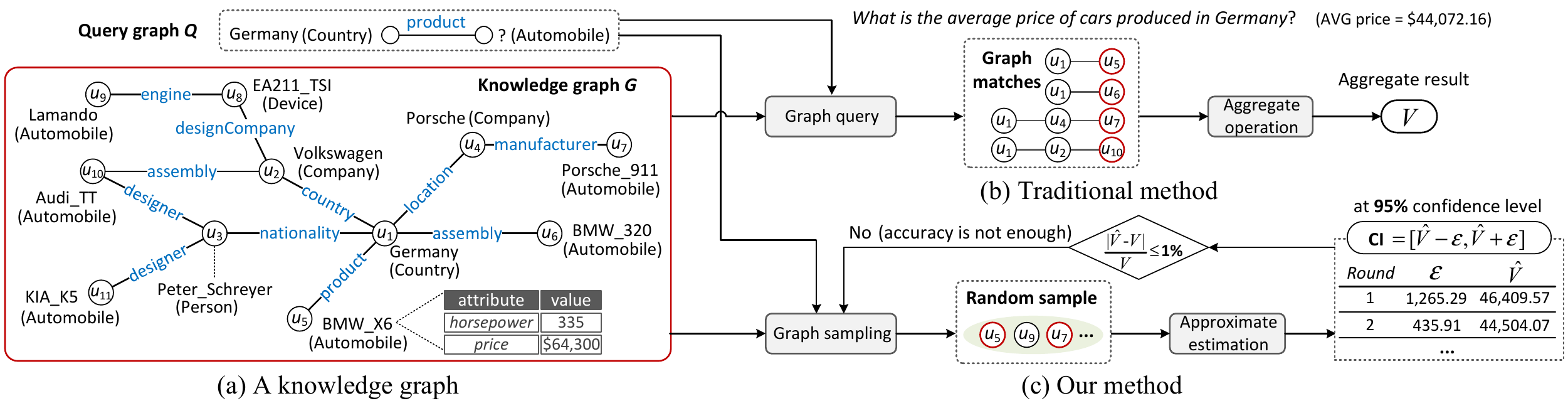}}
\caption{\small (a) Each entity of this knowledge graph with type \textit{Automobile} has many numerical attributes, including \textit{horsepower}, \textit{price}, etc. (b) The traditional method computes an aggregate result based on the graph matches via graph query. (c) Our method computes an approximate aggregate result with an accuracy guarantee in the form of confidence interval (CI) through a ``sampling-estimation'' model.}
\label{fig:fullexample}
\vspace{-0.6cm}
\end{figure*}

\vspace{0.1cm}
\noindent{\textbf{Factoid query}.} 
The answers to a factoid query are defined as an enumeration of noun phrases \cite{Agichtein2005}, e.g., ``\textit{Find all cars produced in Germany}'' (Q117 from QALD-4 benchmark \cite{qald}). Given the KG in Figure \ref{fig:fullexample}(a),
we expect answers as all entities having type \textit{Automobile} that satisfy the semantic relation \textit{product} to the specific entity \textit{Germany}, e.g., ${\sf Audi\_TT}$ ($u_{10}$), ${\sf BMW\_320}$ ($u_6$), etc. Notice that these correct answers are linked with \textit{Germany}
in structurally different ways in Figure \ref{fig:fullexample}(a), for instance, $u_{10}$: Audi\_TT-\textit{assembly}-Volkswagen-\textit{country}-Germany; $u_{6}$: BMW\_320-\textit{assembly}-Germany. This reflects the ``\textit{schema-flexible}" nature of a KG and we expect to find all the semantically similar answers for factoid queries.

\vspace{0.1cm}
\noindent{\textbf{Aggregate query}.}  A simple aggregate query is used to explore the statistical result of a set of entities given a specific entity and a semantic relation. For example, ``\textit{what is the average price of cars produced in Germany?}'' is an aggregate query to achieve ${\sf AVG(price)}$ of all the \textit{Automobiles} that satisfy the semantic relation \textit{product} to the specific entity \textit{Germany}.
We find that 31\% queries from the real query log \textit{LinkedGeoData13} and 30\% queries from the manually curated query set \textit{WikiData17} are aggregate queries \cite{Bonifati2017}.

One frequently used technology to answer factoid queries is graph query \cite{Khan2013,Zou2014,Yang2016,Jin2015,Zou2011,Wang2020}, which we adopt in this work: A user constructs a query graph $Q$ to describe her query intention, and identifies the exact or approximate matches of $Q$ in a KG $G$. 
We can also reduce other query forms, such as keywords and natural languages \cite{Yang2016}, to graph queries by translating input text to a query graph \cite{Zheng2015a,Han2017}. In contrast, answering aggregate queries on KGs has been mostly ignored in the literature. Aggregate queries can be extended from factoid queries, by applying an additional aggregation on factoid queries' answers to obtain the statistical result of interest \cite{HuDYY18,Zou2011} (as Figure \ref{fig:fullexample}(b) shows). However, this straightforward method is problematic due to following reasons.

\vspace{0.1cm}
\noindent\textit{Effectiveness issue.} If aggregate queries are answered via factoid queries, its effectiveness would depend on the quality of factoid queries' returned answers. For example, subgraph isomorphism \cite{Cheng2008,Zou2011} only returns answers that {\em exactly} match with the given query graph $Q$ (e.g., only $u_5$ is returned for $Q$ in Figure \ref{fig:fullexample}), while other semantically similar but structurally different answers are ignored (e.g., $u_6$, $u_7$, and $u_{10}$). Analogously, a relational or SPARQL query finds answers matching {\em exactly} the schema of the input query, and
other valid answers with different schemas will be ignored (see \cite{Khan2011,Wang2020,Yang2014} and also our experimental results).
In addition to exact matching, several other works \cite{Khan2013,Yang2016,Fan2010a,Zheng2016} return similar answers to $Q$. However, it is difficult for them to return 100\% accurate answers (the notion of ``accurate'' answers could very well depend on the user's query intension, or may even be vague \cite{LPHM20,WuK19}). Calculating the aggregate result over answers with low quality leads to significant errors. Worse still, we lack an effective way to quantify the result's quality.

\noindent\textit{Efficiency issue.} Since an additional aggregate operation is applied on the factoid queries' answers to obtain the aggregate result, factoid queries’ efficiency substantially affects aggregate queries’ efficiency. Finding answers to a given query graph $Q$, however, is computationally expensive (e.g., tens of seconds are required in \cite{Fan2010a}). Even the top-$k$ graph query models still need hundreds of milliseconds to tens of seconds to respond \cite{Jin2014, Yang2016, Zheng2016, Wang2020, Zou2011}.

In practice, aggregate queries may not need a tardy exact result. 
It is more desirable if a query engine first quickly returns an approximate aggregate result with some accuracy guarantee (e.g., a confidence interval), while improving the accuracy as more time is spent \cite{Laptev2012,ChaudhuriDK17}. In this way, we can early terminate the query once the approximate result is acceptable. This improves the user's experience and saves computing resources \cite{Li2016,Wang2020,Bhowmick2013,AgarwalMPMMS13}.

\vspace{0.1cm}
\noindent{\textbf{Our solution}.}
Due to the ``\textit{schema-flexible}'' nature in KGs, we adopt the ``\textit{semantic similarity}'' \cite{Wang2020} (defined in ${\rm\S \ref{exact}}$) to measure how semantically similar a candidate answer is to a query graph. We then propose an iterative and approximate approach to efficiently answer aggregate queries over KGs, having an accuracy guarantee, but without requiring factoid query evaluations. As Figure \ref{fig:fullexample}(c) shows, we first collect answers that are semantically similar to a query graph $Q$ as a random sample from a KG $G$ (\textbf{\S \ref{sampling}}). Next, we estimate an unbiased (or consistent) approximate aggregate result $\hat{V}$ based on the random sample (\textbf{\S \ref{estimatoraccurancy}}), and provide an accuracy guarantee for $\hat{V}$ by iteratively computing a tight enough confidence interval CI $=[\hat{V}-\varepsilon,\hat{V}+\varepsilon]$ at a confidence level $1-\alpha$, where $\varepsilon$ is the half-width of a CI (also called the Margin of Error). A CI states that the ground truth $V$ is covered by an interval $\hat{V}\pm \varepsilon$ with probability $1-\alpha$. We terminate the query when a tight CI (with a small enough $\varepsilon$) is obtained, and ensure that the relative error of $\hat{V}$ is bounded by a user-specific error bound $e_b$ (\textbf{\S \ref{accuracy}}). To the best of our knowledge, \textit{we are the first to use a ``\textit{sampling-estimation}'' model to answer aggregate queries on KGs with an accuracy guarantee, together with iterative improvements in error bounds}. 

Given a query graph $Q$, it is \textit{non-trivial to collect answers that are semantically similar to $Q$ as a random sample from a KG $G$}. First, we cannot directly apply the sampling approaches for relational datasets \cite{Li2016, Wu2010,Hellerstein1997,Mozafari2015,HuangYPM19,ChaudhuriDN07} in our case, because a KG’s structure differs from that of a relational data. Though we can model graphs as relations, it adds overhead in the query: We need expensive joins to generate intermediate views and then sample and aggregate over them. Furthermore, \textit{relational or SPARQL query would not find those valid answers having different schemas from the input query} \cite{Khan2011,Wang2020,Yang2014}. Second, existing graph sampling approaches, including CNARW and Node2Vec \cite{Li2019,Li2015,Chiericetti2016,Leskovec2006,Grover2016}, \textit{only consider topology information for sampling, which ignores semantic information in a KG}, therefore an answer in the random sample could probably have a low semantic similarity to $Q$.

To this end, we leverage an offline KG embedding model \cite{Bordes2013,Ji2015,Wang2014} to represent predicates as $d$-dimensional vectors that can well capture their semantic meanings and measure the predicate similarity. On top of this, we design a semantic-aware sampling algorithm via a random walk on $G$. As a result, answers that are more semantically similar to $Q$ would be sampled with higher probabilities than others with lower semantic similarities. We formally prove that the random walk converges, and all answers in a random sample are independent and identically distributed (i.i.d.)
random variables.

\vspace{0.1cm}
\noindent{\textbf{Contributions}.}
Our key contributions include {\bf (1)} \textit{designing of a random walk following predicate similarity via KG embedding} to collect a high-quality sample of answers which are semantically similar to the query graph, {\bf (2)} \textit{theoretical characterization of our random walk and proposed estimators} that answer aggregate queries over KGs with accuracy guarantees and iterative improvements, {\bf (3)} \textit{extending our solution to support complex queries} with filter, GROUP-BY, and different graph shapes, e.g., chain, cycle, star, and flower \cite{Bonifati2017} (\textbf{\S \ref{general}}), and {\bf (4)} \textit{thorough experiments over three diverse real-world KGs} showing accuracy and efficiency improvements against state-of-the-art methods, and our approach's effectiveness when a user varies the error bound interactively (\textbf{\S \ref{experiment}}).

As the first step, we mainly focus on non-extreme aggregates $\{{\sf COUNT,SUM,AVG}\}$ with accuracy guarantees. Notice that our solution can also support extreme functions, e.g., ${\sf MAX}$, ${\sf MIN}$ without accuracy guarantees (see \S \ref{experiment}), while in future we will study their theoretical accuracy guarantees. Related work is discussed in \textbf{\S \ref{previous_work}}, while in \textbf{\S \ref{conclusion}} we conclude. 

\section{Preliminaries}
\label{pre_overview}
We first provide the preliminaries and then formalize the problem that we study in this paper. Frequently used notations are summarized in Table \ref{tab:notations}.
\begin{myDef}
\label{def:kg}
\textbf{Knowledge graph (KG)}. A KG is defined as $G=(V_G,E_G,L_G,A_G)$, where $V_G$ is a finite set of nodes and $E_G\subseteq V_G\times V_G$ is a set of edges. (1) Each node $u\in V_G$ represents an entity and each edge $e\in E_G$ denotes a relationship between two entities. (2) A label function $L_G$ assigns a name and various types on each node $u\in V_G$, and  a predicate on each edge $e\in E_G$. (3) Each node $u\in V_G$ has a set of numerical attributes, denoted by $A_G(u)=\{a_1...a_n\}$, and $u.a_i$ indicates the value of attribute $a_i$ of $u$.
\end{myDef}
%

We assume that each node $u$ in a KG $G$ has at least one type and a unique name \cite{Zheng2016, Zheng2018}. If the node type is unknown, we employ a probabilistic model-based entity typing method to assign a type on it \cite{Nakashole2013}. For example, $L_G(u).{\sf type}=\{{\sf Automobile, MeanOfTransportation}\}$; $L_G(u).{\sf name}={\sf BMW\_X6}$. For each edge $e$, it has a predicate such as $L_G(e)={\sf assembly}$. Moreover, each node $u$ has a set of numerical attributes, e.g., $A_G(u)=\{{\sf horsepower, price, \cdots}\}$, such as $u.{\sf horsepower}=335$ for ${\sf BMW\_X6}$.
%

\begin{myDef}
\label{def:ag}
\textbf{Aggregate query over $G$}. An aggregate query over $G$ is defined as $AQ_G=(Q,f_a)$, where $Q$ is a query graph for searching candidate answers from $G$ and $f_a$ is an aggregate function on the numerical attribute $a$ of the answers to $Q$. In this paper, we primarily consider three widely-used non-extreme aggregate functions $\{{\sf COUNT,SUM,AVG}\}$.
\end{myDef}

\begin{figure}
\vspace{-0.3cm}
\setlength{\abovecaptionskip}{0.1cm}
\centerline{\includegraphics[scale=0.6]{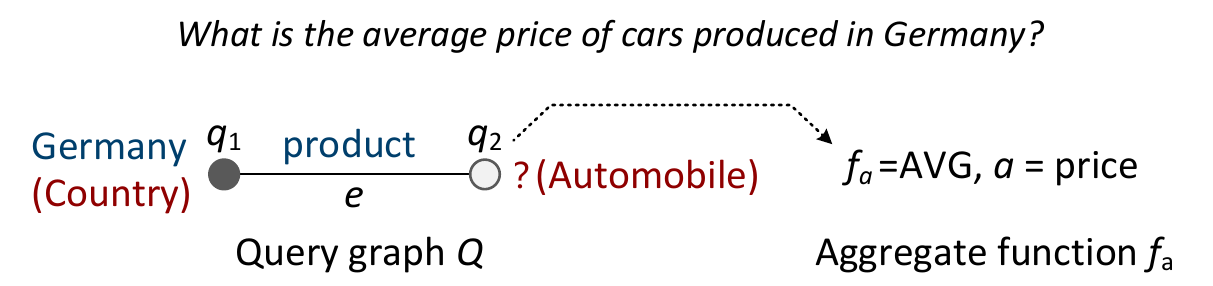}}
\caption{\small An example of aggregate query $AQ_G=(Q,f_a)$}
\label{fig:AQG}
\vspace{-0.6cm}
\end{figure}
%

We start with the query graph $Q$ for simple questions --- one of the most common questions \cite{Bao2016,Bordes2015}, involving a single specific entity and a single predicate, taking the target entities as the answers \cite{Huang2019}. Given our proposed framework to answer $AQ_G=(Q,f_a)$ for simple questions, we use it as a building block to support more general cases (discussed in \S \ref{general}). 
%

\begin{myDef}
\label{def:qg}
\textbf{Query graph}. A query graph is defined as a graph $Q=(V_Q,E_Q,L_Q)$, with query node set $V_Q$, edge set $E_Q$, and label function $L_Q$. For a simple question,$V_Q=\{q^s,q^t\}$ contains two nodes, where $q^s$ is a specific node and $q^t$ is a target node. For $q^s$, both the types and name are known, while for $q^t$, only the types are known. Moreover, $E_Q$ has one edge $e=q^sq^t$ with a predicate $L_Q(e)$.
\end{myDef}
%

\begin{myExample}
\label{exp:AQG}
Given a simple question ``{\em what is the average price of cars produced in Germany?}'', we formulate $AQ_G=(Q,f_a)$ in Figure \ref{fig:AQG}. The query graph $Q$ contains a specific node $q^s=q_1$ $($type: $\{{\sf Country}\}$, name: ${\sf Germany})$, a target node $q^t=q_2$ $($type: $\{{\sf Automobile}\})$, an edge $e=q_1q_2$ $($predicate: ${\sf product})$, and $f_a={\sf AVG}$ on the attribute $a={\sf price}$.
\end{myExample}
%

\begin{myDef}
\label{def:answers}
\textbf{Candidate answers to $Q$}. Given a query graph $Q$ and a KG $G$, candidate answers $\mathcal{A}=\{u^t_1,\cdots, u^t_n\}$
are certain nodes from $G$: (1) Each $u^t_i$ must have at least one common type as the target node $q^t$ from $Q$ (i.e., $L_G(u^t_i).{\sf type}\cap L_Q(q^t).{\sf type}\neq \emptyset$). (2) Each $u^t_i$ has semantic similarity $s_i\in [0,1]$ (defined in ${\rm\S \ref{exact}}$), indicating how semantically similar the best subgraph match containing $u^t_i$ is to $Q$.
\end{myDef}

\vspace{0.05cm}
Due to the schema-flexible nature of KGs, the same kind of information can be represented as different substructures \cite{Zheng2016, Wang2020}. So we find many semantically similar but structurally different subgraph matches to a given query graph $Q$. This motivates us to adopt the semantic similarity to measure how semantically similar a candidate answer is to $Q$. We introduce a tunable parameter $\tau$ as the threshold and view those candidate answers having $s_i\geq \tau$ as the {\em correct answers} to $Q$, denoted by $\mathcal{A}^+=\{u^t_i\in \mathcal{A}:s_i\geq \tau\}$.
\begin{table}
  \centering
  \setlength{\abovecaptionskip}{0.1cm}
  \footnotesize
  \caption{\small Frequently used notations}
  \label{tab:notations}
  \begin{tabular}{c||l}
    \textbf{Notations} & \textbf{Descriptions}\\
    \hline
    \hline
    $G$ & A knowledge graph\\
    \hline
    $AQ_G=(Q,f_a)$ & \makecell[l]{An aggregate query over $G$ with a query graph $Q$\\and an aggregate function $f_a$} \\
    \hline
    $\mathcal{A}$ &  A set of candidate answers to $Q$;  $\mathcal{A}=\{u^t_1,\cdots, u^t_n\}$ \\
    \hline
    $S_\mathcal{A}$ & A random sample of answers collected from $\mathcal{A}$\\
    \hline
    $s_i$ & The semantic similarity of an answer $u^t_i\in \mathcal{A}$ to $Q$\\
    \hline
    $\tau$ & A user-input semantic similarity threshold \\
    \hline
    $\mathcal{A}^+\subseteq \mathcal{A}$ & A set of correct answers to $Q$; $\mathcal{A}^+=\{u^t_i\in \mathcal{A}:s_i\geq \tau\}$ \\
    \hline
    $V$ & The ground truth of $AQ_G$; $V$=$f_a(\mathcal{A}^+)$ \\
    \hline
    $\hat{f}_a$ & An unbiased (or consistent) estimator of $f_a$ \\
    \hline
    $\hat{V}$ & The estimated approximate result of $AQ_G$; $\hat{V}$=$\hat{f}_a(S_\mathcal{A})$ \\
    \hline
    $e_b$ & A user-input error bound \\
    \hline
    $1-\alpha$ &  A user-input confidence level\\
    \hline
    $\hat{V}\pm \varepsilon$ &  The confidence interval (CI) at $1-\alpha$ confidence level\\
    \hline
    $\varepsilon$ &  The half width of CI, called the Margin of Error (MoE)\\
  \end{tabular}
  \vspace{-0.1cm}
\end{table}

\begin{table}
\vspace{-0.2cm}
\setlength{\abovecaptionskip}{0.1cm}
  \caption{\small Some candidate answers to the query ($Q$) in Figure \ref{fig:AQG}}
  \centerline{\includegraphics[scale=0.4]{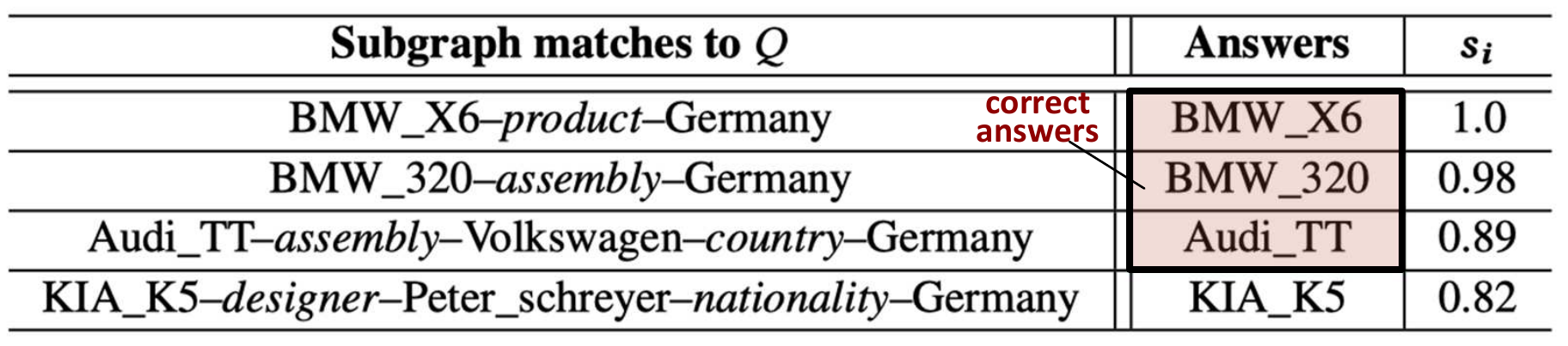}}
\label{tab:answers}
\vspace{-0.7cm}
\end{table}
%

\begin{myExample}
\label{cananswers}
Table \ref{tab:answers} shows four candidate answers to the query graph $Q$ in Figure \ref{fig:AQG}. Each answer (second column) is an entity from the subgraph match to $Q$ (first column) that has a semantic similarity to $Q$ (third column), e.g., ${\sf BMW\_X6}$ (first row) has a semantic similarity of $1.0$ because its subgraph match is exactly the same as $Q$. Also, ${\sf KIA\_K5}$ (last row) has a semantic similarity of $0.82$ because its subgraph match is semantically quite different from $Q$. By setting $\tau=0.85$, ${\sf KIA\_K5}$ can be eliminated from the correct answers.
\end{myExample}
%

A domain expert can tune $\tau$ appropriately according to her experience and based on available human annotation
(an example is given in \S \ref{experiment}). So, we resort to semantic similarity-based ground truth
(or $\tau$-relevant ground truth, abbreviated as $\tau$-GT) for evaluation, in addition to human annotation-based ground truth
(abbreviated as HA-GT) when available. 
We obtain $\tau$-GT $V$ of $AQ_G$=$(Q,f_a)$ by applying $f_a$
on all those $\tau$-relevant correct answers $\mathcal{A}^+$ to $Q$. That is, $V$=$f_a(\mathcal{A}^+)$.
Given above definitions, we are interested in the following problem.

\vspace{0.05cm}
\noindent\underline{\textbf{Approx-$\mathbf{AQ_G}$.}} Given an aggregate query $AQ_G=(Q,f_a)$, a KG $G$, an input error bound $e_b$, and a confidence level $1-\alpha$, we aim to: (1) design a sampling algorithm $\mathcal{D}$ to collect a random sample $S_\mathcal{A}=\mathcal{D}(\mathcal{A})$ of the candidate answers $\mathcal{A}$ from $G$, (2) estimate the approximate result $\hat{V}$ based on $S_\mathcal{A}$ with a confidence interval $\hat{V}\pm \varepsilon$ at $1-\alpha$ confidence level, and (3) ensure that the relative error of $\hat{V}$ is bounded by $e_b$.
\begin{equation}
\small
\label{eq:approx}
\vspace{-1.5mm}
\hat{V}=\hat{f}_a(S_\mathcal{A}) \nonumber
\end{equation}
\vspace{-3mm}
\begin{equation}
\small
\label{eq:approx1}
s.t. \quad \Pr[\hat{V}-\varepsilon\leq V\leq \hat{V}+\varepsilon]=1-\alpha\quad \text{and} \quad |\hat{V}-V|/V\leq e_b
\vspace{-0.8mm}
\end{equation}

In Eq. \ref{eq:approx}, $\hat{f}_a$ is an unbiased (or consistent) estimator of the given aggregate function $f_a$. The approximate result $\hat{V}$ is a point estimator to the ground truth $V$ and a $1-\alpha$ level confidence interval CI $=[\hat{V}-\varepsilon,\hat{V}+\varepsilon]$ is computed to quantify the point estimator's quality, which states that $V$ is covered by an estimated range $\hat{V}\pm\varepsilon$ with probability $1-\alpha$. The half width of CI, denoted by $\varepsilon$, is called the Margin of Error (MoE). Generally, the smaller MoE shows the higher quality of $\hat{V}$, i.e., $\hat{V}$ is much closer to $V$. In \S \ref{accuracy}, we prove that the accuracy guarantee $|\hat{V}-V|/V\leq e_b$ is ensured if the MoE is small enough to satisfy $\varepsilon\leq \hat{V}\cdot e_b/(1+e_b)$ (Theorem \ref{th:accuracy}). Otherwise, we enlarge the sample $S_\mathcal{A}=S_\mathcal{A}\cup\Delta S_\mathcal{A}$ and repeat Eq. \ref{eq:approx} to continuously refine the CI $=\hat{V}\pm \varepsilon$ until $\varepsilon\leq \hat{V}\cdot e_b/(1+e_b)$.

\vspace{0.1cm}
\noindent \underline{\textbf{Remarks.}} In our empirical study (\S \ref{experiment}), we show that our solution to Approx-$\rm AQ_G$ (proposed in \S \ref{ours}) has a good accuracy w.r.t. $\tau$-GT. Next, we demonstrate that our solution can also have a good accuracy w.r.t. HA-GT by setting an appropriate $\tau$, which ensures that $\tau$-GT and HA-GT are highly similar. Thus, in the absence of HA-GT in practice, domain experts may still set a good $\tau$ to achieve an acceptable solution quality.

%
\section{The Semantic Similarity Baseline}
\label{exact}
Before discussing our approximate solution, we introduce a simple, but costly enumeration method, called
\textit{\textbf{S}emantic \textbf{S}imilarity-based \textbf{B}aseline} (\texttt{SSB}), to answer an aggregate query $AQ_G$=$(Q,f_a)$.
We apply \texttt{SSB} to get semantic similarity-based ground truth ($\tau$-GT) for effectiveness evaluation (\S \ref{experiment}). 

\setlength{\textfloatsep}{0cm}
\setlength{\intextsep}{0cm}
\begin{algorithm}[t]
\footnotesize
\caption{{\small Semantic Similarity-based Baseline (\texttt{SSB})}}
\label{alg:exact_alg}
\KwData{knowledge graph $G$, aggregate query $AQ_G=(Q,f_a)$, threshold $\tau$, and subgraph boundary $n$}
\KwResult{aggregate result $V$ of the $\tau$-relevant correct answers}
\tcp{\small{$n$-bounded subgraph construction}}
$u^s=$ getMappingNode($Q.q^s$)\;
$G'=$ getBoundedGraph($G$, $u^s$, $n$)\;
\tcp{\small{$\tau$-relevant correct answers enumeration}}
\For{$\forall u^t_i\in \mathcal{A}\subseteq G'$}{
	$s_i=$ getSimilarity($u^t_i$,$Q$)$\geq \tau$ : $\mathcal{A}^+$.add($u^t_i$) ? continue\;
}
\Return $V=f_a(\mathcal{A}^+)$\;
\end{algorithm}

The basic idea of \texttt{SSB} (Algorithm~\ref{alg:exact_alg}) is to enumerate all candidate answers $\mathcal{A}$, find all correct answers $\mathcal{A}^+\subseteq\mathcal{A}$ having semantic similarities $s_i\geq\tau$ ($\tau$ is a predefined threshold), and then compute the aggregate result over $\mathcal{A}^+$. Considering all candidate answers, however, is unnecessary, because graph queries exhibit strong access locality \cite{Yang2012}, thus most correct answers could be found in an $n$-bounded space of the specific node \cite{Wang2020} (in \S \ref{experiment}, we empirically find that $n$=3 can retrieve 99\% of all correct answers). Hence, it is reasonable to limit the search space of \texttt{SSB} in a $n$-bounded subgraph $G'$ of $G$.

\vspace{0.05cm}
\noindent\underline{\textbf{$n$-bounded subgraph construction.}} Given a query graph $Q$ and a KG $G$, we get the mapping node $u^s$ from $G$ for the specific node $q^s$ from $Q$ that satisfies: $L_G(u^s).{\sf name}=L_Q(q^s).{\sf name}$, $L_G(u^s).{\sf type}\cap L_Q(q^s).{\sf type}\neq \emptyset$ (Line 1). Then we conduct a BFS starting from $u^s$ to construct the $n$-bounded subgraph $G'$, of which each entity $u$ from $G'$ is within $n$-hops from $u^s$ (Line 2). Since a KG adopts some entity disambiguation methods \cite{Cucerzan2007,Li2016a,Hu2020} to ensure that each node has a unique name, for simplicity we assume that $q^s$ has a unique mapping node $u^s$. We next introduce how to measure the semantic similarity of each candidate answer from $G'$ (Lines 3-4). This is the most expensive step in \texttt{SSB}.

\vspace{0.05cm}
\noindent\underline{\textbf{Semantic similarity of an answer.}} We start with defining a subgraph match $M(u^t_i)$ of a candidate answer $u^t_i\in \mathcal{A}$. 

\vspace{0.05cm}
\begin{myDef}
\label{def:exactmatch}
\textbf{Subgraph match} \cite{Wang2020}. Given a simple query graph $Q$ and an $n$-bounded subgraph $G'$, a subgraph match $M(u^t_i)$ to $Q$ is defined as an edge-to-path mapping from the query edge $e=q^sq^t$ in $Q$ to a path $\overline{u^su^t_i}$ in $G'$. (1) For the specific node $q^s$, $u^s$ is its mapping node satisfying $L_G(u^s).{\sf name}=L_Q(q^s).{\sf name}$ and $L_G(u^s).{\sf type}\cap L_Q(q^s).{\sf type}\neq \emptyset$. (2) For the target node $q^t$, $u^t_i$ is a candidate answer that satisfies $L_G(u^t_i).{\sf type}\cap L_Q(q^t).{\sf type}\neq \emptyset$.
\end{myDef}
%

For example, in Table \ref{tab:answers}, the subgraph match $\langle{\sf Audi\_TT}$-\textit{assembly}-${\sf Volkswagen}$-\textit{country}-${\sf Germany}\rangle$ contains an answer ${\sf Audi\_TT}$ (i.e., $u^t_i$). 
%
%
Intuitively, a subgraph match $M(u^t_i)$ is more semantically similar to $Q$ if each edge $e'$ on the path $\overline{u^su^t_i}$ is more semantically similar to the query edge $e$ in $Q$. Following \cite{Wang2020}, we define the semantic similarity $s[M(u^t_i)]$ of $M(u^t_i)$ to $Q$ as the geometric mean of the predicate similarities of all edges in $\overline{u^su^t_i}$ (Eq. \ref{eq:pss}), where $sim(L_G(e')$,$L_Q(e))$ is the predicate similarity between $e'$ and $e$; $l$ is the length of $\overline{u^su^t_i}$.  If there are multiple subgraph matches of $u^t_i$, we compute the semantic similarity $s_i$ of $u^t_i$ as the maximum semantic similarity considering all its subgraph matches (Eq. \ref{eq:sss}).
\begin{equation}
\small
s[M(u^t_i)]=\sqrt[\leftroot{2}\uproot{4} l]{\prod_{e'\in \overline{u^su^t_i}}{sim(L_G(e'),L_Q(e))}}
\label{eq:pss}
\end{equation}
\vspace{-0.25cm}
\begin{equation}
\small
s_i=\max_{M(u^t_i)}s[M(u^t_i)]
\label{eq:sss}
\end{equation}
%
%

We leverage an offline {\em KG embedding} model to obtain the predicate similarity between two edges $e'$ and $e$. A KG embedding aims to represent each predicate and entity in a KG $G$ as a $d$-dimensional vector, it can preserve well the semantic meanings and relations using these learned semantic vectors \cite{Huang2019}. We refer interested readers to \cite{Bordes2013,Wang2020} for more details. The similarity between two predicates $sim(L_G(e'),L_Q(e))$ (e.g., $sim({\sf assembly}$, ${\sf product})$) can be computed by the cosine similarity between their predicate vectors $\bm{e}$ and $\bm{e'}$.
\begin{equation}
\small
\label{eq:cosine}
sim(L_G(e'),L_Q(e))=\frac{\bm{e'}\cdot \bm{e}}{||\bm{e'}||\times||\bm{e}||}
\end{equation}
%

\begin{myExample}
\label{exp:pss}
Consider the answer ${\sf Audi\_TT}$ in Table \ref{tab:answers} for the query graph in Figure \ref{fig:AQG}. Its semantic similarity is $\sqrt[2]{0.98\times 0.81}$=$0.89$, where $sim({\sf assembly}$, ${\sf product})$=$0.98$ and $sim({\sf country}$, ${\sf product})$=$0.81$, based on the TransE model \cite{Bordes2013}.
\end{myExample}

In \S \ref{experiment}, we apply \texttt{SSB} to get semantic similarity-based ground truth (i.e., $\tau$-relevant ground truth, $\tau$-GT) for effectiveness evaluation, besides human annotation-based ground truth (HA-GT). In fact, \texttt{SSB} can work with any underlying KG embedding. Ideally, if we have a high-quality KG embedding model, then we can distinguish the implicit semantics of predicates well by Eq. \ref{eq:cosine}; hence, we can also effectively represent the semantics of different paths and answers by Eq. \ref{eq:pss}-\ref{eq:sss}. So, it is likely that both $\tau$-GT and HA-GT would be similar for an appropriate $\tau$. In \S \ref{experiment}, we present the effectiveness of our approximate solution (\S \ref{ours}) w.r.t. both $\tau$-GT and HA-GT. We also study the effect of KG embedding models on effectiveness.

\vspace{0.05cm}
\noindent\underline{\textbf{Remarks.}} (1) Different from structure-based similarity which assumes that a shorter path indicates higher similarity \cite{Khan2013,Jin2017,Jin2014}, our semantic similarity captures the implicit semantics of a path through KG embedding. In fact, the path length usually does not reflect the semantics of a path, and a longer path might have a higher semantic similarity than a shorter one to a given query graph. This is realistic, for instance, a longer path (BMW\_Z4-\textit{assembly}-Regensburg-\textit{federalState}-Bavaria-\textit{country}-Germany) may be semantically more similar than a shorter path (KIA\_K5-\textit{designer}-Peter\_schreyer-\textit{nationality}-Germay) w.r.t. the query graph in Figure~\ref{fig:AQG}. (2) As semantic similarity of a path is non-monotonic w.r.t. its length (Eq.~\ref{eq:pss}), Dijkstra-like algorithm is inadequate to directly find the path from $u^s$ to $u^t_i$ with the highest semantic similarity. Instead, one needs to enumerate all paths from $u^s$ to $u^t_i$, compute their semantic similarities (Eq.~\ref{eq:pss}), and thus find the best one (Eq.~\ref{eq:sss}).
(3) The time complexity of \texttt{SSB} is: $O(|\mathcal{A}|\cdot m^n)$, where $m$ is the average degree of an entity in a KG and $m^n$ is the search space of path enumeration for each candidate answer in $\mathcal{A}$. Note that in our implementation, we do not explicitly construct the $n$-bounded subgraph $G'$, rather all paths up to length $n$
from $u^s$ are considered for each candidate answer in $\mathcal{A}$. Due to the inefficiency of \texttt{SSB}, we introduce a lightweight ``sampling-estimation'' solution in \S \ref{ours}. (4) Following \cite{Wang2020}, we use geometric mean to define semantic similarity of a subgraph match (Eq.~\ref{eq:pss}). This is one of the best-efforts to define correct answers in the absence of human-annotated ground truth. Interestingly, our ``sampling-estimation’’ method (\S \ref{ours}) does not strictly depend on Eq.~\ref{eq:pss}, instead it works well so long as the semantic similarity of a subgraph match remains monotone w.r.t. predicate similarity between each edge in the subgraph and the query edge. We select a higher threshold $\tau$ to avoid lower-quality matches.


\section{Sampling-Estimation Solution}
\label{ours}
We first provide a high-level introduction to our approximate ``sampling-estimation" solution, then we drill down into the details (\S \ref{sampling}-\ref{accuracy}) and show the entire algorithm in \S \ref{alg_blk}. Given a KG $G$ and an aggregate query $AQ_G=(Q,f_a)$, our solution consists of three steps. (1) \textit{Semantic-aware sampling on KGs} (\S \ref{sampling}): We collect answers with higher semantic similarities to $Q$ as a random sample $S_\mathcal{A}$, through a semantic-aware random walk sampling over $G$. (2) \textit{Approximate result estimation} (\S \ref{estimatoraccurancy}): We consider the semantic similarity to design unbiased estimators for $\{{\sf COUNT}$, ${\sf SUM}\}$ and a consistent estimator for ${\sf AVG}$, then we apply them on $S_\mathcal{A}$ to estimate the approximate result $\hat{V}$. (3) \textit{Accuracy guarantee} (\S \ref{accuracy}): We derive an accuracy guarantee based on the \textit{Central Limit Theorem} (CLT) in the form of confidence interval CI $=\hat{V}\pm \varepsilon$ and iteratively refine the CI until an acceptable accuracy (Theorem \ref{th:accuracy}) is attained.

%
%

\vspace{-0.2cm}
\subsection{Semantic-aware Sampling on KGs}
\label{sampling}
%
\subsubsection{Classic Random Walk on Graphs}
\label{rws}
Random walk is the mainstream technique for graph sampling due to its scalability and simplicity of implementation \cite{Zhao2015,Li2019}. A general random walk sampling on a KG $G$ can be modeled as a finite Markov Chain \cite{Li2019},  having the following steps. A walker starts from a randomly selected node, say $u_0\in V_G$, then randomly chooses a neighbor of $u_0$ and moves to it with the transition probability defined in the transition matrix $\bm{P}=|V_G|\times |V_G|$. This walker continues to walk until a stationary distribution $\bm{\pi}=\{\pi_1,\cdots,\pi_{|V_G|}\}$ is reached, where $\sum \pi_i=1$ and $\pi_i$ is the stationary visiting probability of each node $u_i\in V_G$ when the random walk converges. The walker keeps walking after the stationary distribution $\bm{\pi}$ is reached and collects all visited nodes as a random sample of $V_G$. Each collected node $u_i$ can be viewed as being sampled with its visiting probability $\pi_i$.

The transition matrix $\bm{P}$ is the key to the random walk sampling. Different $\bm{P}$ are required for different downstream applications. Unlike previous topology-aware graph sampling works \cite{Grover2016,Zhao2015,Li2019}, \textit{we are the first to develop a semantic-aware graph sampling on KGs for aggregate queries}.

\subsubsection{Semantic-aware Random Walk Sampling}
\label{srws}
Similar to \texttt{SSB}, we expect to run our semantic-aware random walk sampling on the $n$-bounded subgraph $G'$ rather than on the entire $G$, to improve the efficiency by reducing the scope of random walk. To achieve this, in our implementation, we limit the random walk on $G$ within $n$-hops from the mapping node $u^s$. To be precise, a node $u$ is within $n$-hops from $u^s$ if there is at least one path with length $\leq n$ between them. For simplicity of understanding, we still use $G'$ (the $n$-bounded subgraph) to indicate the scope of our random walk, this is the induced graph formed by those nodes within $n$-hops from $u^s$. We next introduce our \textit{design of transition matrix} $\bm{P}$ by considering the semantic similarity. Then, we discuss two phases of our semantic-aware sampling based on $\bm{P}$: \textit{random walk until convergence} and \textit{continuous sampling}.


\vspace{0.05cm}
\noindent\underline{\textbf{(1) Design of transition matrix.}} With the $n$-bounded subgraph $G'$, we design a $|V_{G'}|\times |V_{G'}|$ transition matrix $\bm{P}=[p_{ij}]$, where $V_{G'}$ is the node set of $G'$. Since we expect to collect answers having higher semantic similarities as a random sample $S_\mathcal{A}$, we ensure that such an answer has a greater visiting probability $\pi_i$ when the random walk converges to a stationary distribution $\bm{\pi}$. The greater $\pi_i$ is, the higher is the probability of this answer being sampled.

\begin{figure}
\vspace{-0.4cm}
\setlength{\abovecaptionskip}{0.1cm}
\centerline{\includegraphics[scale=0.5]{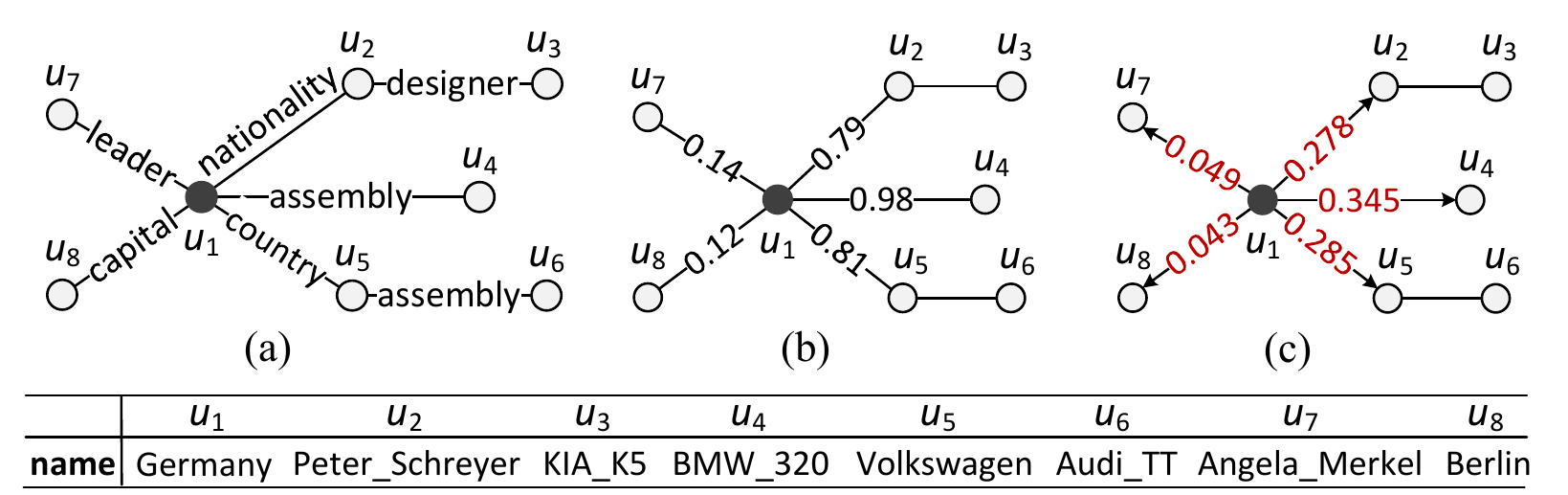}}
\caption{\small (a) A KG with a table of the entities' names. (b) The predicate similarities of $u_1$'s adjacency edges to the predicate $\sf{product}$. (c) The transition probabilities $p_{1j}$ for $u_j\in N(u_1)$.}
\label{fig:transition}
\end{figure}

\vspace{0.05cm}
\noindent\textbf{Transition probability.} Intuitively, if we can design a transition matrix $\bm{P}$ to guide a walker towards an answer $u^t_i$ with the greatest $s_i$ along the path $\overline{u^su^t_i}$ as much as possible, then this $u^t_i$ is more likely to have a greater visiting probability $\pi_i$ when the random walk converges. To achieve this, we assign a greater transition probability $p_{ij}$ on each edge $e'$ from $G'$ that has a greater predicate similarity $sim(L_G(e'),L_Q(e))$ to the query edge $e$ from $Q$, while assigning a smaller $p_{ij}$ on each edge $e'$ having a smaller $sim(L_G(e'),L_Q(e))$.

Given a node $u_i$ from the $n$-bounded subgraph $G'$, its neighbors are denoted as $N(u_i)$. We use $e'=u_iu_j$ to indicate the adjacency edge between $u_i$ and its neighbor $u_j\in N(u_i)$. We can easily compute the predicate similarity $sim(L_G(e'),L_Q(e))$ by Eq. \ref{eq:cosine} and assign a transition probability $p_{ij}$ on $e'$ that is proportional to $sim(L_G(e'),L_Q(e))$ by Eq. \ref{eq:tprob}, where $Z$ is the normalization constant and the total probability of moving from $u_i$ to other nodes in $N(u_i)$ should equal to one according to the property of Markov Chain.
\begin{equation}
\small
\label{eq:tprob}
p_{ij}=Z\cdot sim(L_G(e'),L_Q(e))
\end{equation}
\vspace{-0.4cm}
\begin{equation}\nonumber
\small
\label{eq:tprob1}
s.t.\quad \sum_{u_j\in N(u_i)}p_{ij}=1
\end{equation}
%
%
\begin{myExample}
\label{exp:normalize}
Figure \ref{fig:transition} illustrates an example of calculating the transition probabilities for the query $AQ_G$ in Figure \ref{fig:AQG}. Figure \ref{fig:transition}(a) shows a KG in which each node is within 2-hops away from $u_1$ ($\sf{Germany}$). In Figure \ref{fig:transition}(b), each adjacency edge of $u_1$ has a predicate similarity to the predicate ${\sf product}$. The transition probability $p_{1j}$ for $u_j\in N(u_1)$, such as $p_{14}=\frac{0.98}{0.98+0.81+0.79+0.14+0.12}=0.345$, is given in Figure \ref{fig:transition}(c). A walker starting from $u_1$ will likely select $u_4$, $u_5$, or $u_2$ as the next step (they are more relevant to reach automobile nodes), rather than other irrelevant nodes (i.e., $u_7$ and $u_8$). Moreover, the automobile $u_4$ will have a higher probability to be sampled than $u_3$ and $u_6$, because $p_{14}>\{p_{15}, p_{12}\}$.
\end{myExample}

\vspace{0.05cm}
\noindent\textbf{Analysis of the convergence.} We must ensure that the random walk using the transition matrix $\bm{P}$ initialized by Eq. \ref{eq:tprob} can converge to a stationary distribution. A finite Markov Chain (MC) can converge if it is \textit{irreducible} and \textit{aperiodic} \cite{Ross2014}. An MC is irreducible if any two nodes are reachable in finite steps.
%
%
\begin{myLemma}
\label{lemma:1}
Our semantic-aware random walk is irreducible.
\vspace{-2mm}
\end{myLemma}

\begin{IEEEproof}
\label{pf:1}
For any node $u_i$ and its neighbor $u_j\in N(u_i)$, we have a nonzero predicate similarity on each edge $u_iu_j$ so that the transition probability $p_{ij}$ is also $>0$. Thus, for any two nodes, they are reachable from each other in finite steps according to all the nonzero $p_{ij}$ between them. Hence, our semantic-aware random walk is irreducible.
\end{IEEEproof}
%

In an MC, each node $u_i$ has period $k$ if any return to $u_i$ must occur in multiples of $k$ steps, and an MC is aperiodic if it has at least one node having period one \cite{Ross2014}. To satisfy the aperiodic property, we change the structure of $G'$ with a small modification: We add an additional self-loop edge on the mapping node $u^s$ with a small enough predicate similarity $p_{ss}$ on it (we set it as 0.001 in this work). A walker starting from $u^s$ tends to walk outward rather than be stuck at $u^s$ due to this small self-loop predicate similarity, so it has little effect on the convergence time. It is easy to verify that our semantic-aware random walk is aperiodic after this modification.

\begin{myLemma}
\label{lemma:2}
Our semantic-aware random walk is aperiodic.
\end{myLemma}
\begin{IEEEproof}
\label{pf:2}
Because the mapping node $u^s$ has a self-loop edge, a walker can start from $u^s$ and return back to itself in one walk step with a transition probability $p_{ss}$. Hence, $u^s$ has period one and our semantic-aware random walk is aperiodic.
\end{IEEEproof}
%

\vspace{0.05cm}
\noindent\underline{\textbf{(2) Random walk until convergence.}} We start the random walk from the mapping node $u^s$ in the $n$-bounded subgraph $G'$. In each walk step, we adopt a walking-with-rejection policy \cite{Li2019} to determine the next node. Specifically, suppose that a walker is currently at node $u_i$, we select a neighbor node $u_j$ from $N(u_i)$ at random and accept it with probability $p_{ij}$. If $u_j$ is rejected, then we repeat this selection until one node is accepted. After moving from $u_i$ to $u_j$, we update the stationary probability $\pi_j$ by Eq. \ref{eq:balance}. We initialize the stationary distribution as $\bm{\pi}=\{1,0,\cdots,0\}$, where the mapping node $u^s$ has $\pi_s=1$ because we start the random walk from $u^s$.
\begin{equation}
\small
\label{eq:balance}
\pi_j=\sum_{u_i\in N(u_j)}\pi_i\cdot p_{ij}
\end{equation}

\vspace{-0.15cm}
Moreover, Eq. \ref{eq:balance} is a standard way to update $\bm{\pi}$ in the theory of Markov Chain, which guarantees that $\sum \pi_i=1$ holds. The random walk converges if $\bm{\pi}$ is no longer changing (i.e., $\bm{\pi}\cdot \bm{P}=\bm{\pi}$). Finally, we obtain the stationary distribution $\bm{\pi}=\{\pi_1,\cdots, \pi_{|V_{G'}|}\}$, where $V_{G'}$ is the node set of $G'$.

\vspace{0.05cm}
\noindent\underline{\textbf{(3) Continuous sampling.}} As mentioned in \S \ref{rws}, the original continuous sampling will return a random sample of node set $V_{G'}$ by continuous walking on $G'$ after convergence. In our case, however, we expect to collect a random sample $S_\mathcal{A}$ of the candidate answers $\mathcal{A}$ instead of a sample of node set $V_{G'}$. Thus, we make the following changes to the original continuous sampling: (1) We first extract the stationary distribution $\bm{\pi}_{\mathcal{A}}=\{\pi'_1,\cdots,\pi'_{|\mathcal{A}|}\}$ of $\mathcal{A}$ from the stationary distribution $\bm{\pi}=\{\pi_1,\cdots,\pi_{|V_{G'}|}\}$ of $V_{G'}$. For each answer $u^t_i\in \mathcal{A}$, we compute its new visiting probability $\pi'_i=\pi_i/\sum \pi_i$ where $\pi_i\in \bm{\pi}$ is the original visiting probability of $u^t_i$, so that the cumulative probability $\sum \pi'_i=1$ holds for the extracted stationary distribution $\bm{\pi}_\mathcal{A}$, and (2) we conduct the continuous sampling to collect each visited answer $u^t_i\in \mathcal{A}$ to $S_\mathcal{A}$ with its probability of $\pi'_i\in \bm{\pi}_\mathcal{A}$, and ignore all the non-answer nodes.
%
\begin{myTheorem}
\label{th:iid}
All the answers in $S_\mathcal{A}$ are independent and identically distributed (i.i.d.) random variables.
\end{myTheorem}
%

\begin{IEEEproof}
\label{pf:iid}
In continuous sampling, the acceptance decision for each answer $u^t_i$ is made independently. That is, whether $u^t_i$ is sampled depends only on its visiting probability $\pi'_i$ and has no relation with the fact that other answers are or are not being sampled. Moreover, each answer's visiting probability $\pi'_i$ comes from the identical distribution $\bm{\pi}_\mathcal{A}$. Hence, all the answers in $S_\mathcal{A}$ are i.i.d random variables.
\end{IEEEproof}

\vspace{0.05cm}
\noindent\underline{\textbf{Remarks.}} We stop sampling after collecting enough answers. A large or small $|S_\mathcal{A}|$ may lead to over- or under-sampling. We discuss how to configure $|S_\mathcal{A}|$ appropriately in \S \ref{accuracy}. 
\vspace{-0.1cm}
\subsection{Approximate Result Estimation}
\label{estimatoraccurancy}
We present unbiased estimators for $\{\sf{SUM}$, $\sf{COUNT}\}$ and a consistent estimator for $\sf{AVG}$, that is, it converges almost surely to the true expectation.

\subsubsection{Aggregation Estimators}
\label{estimator}
We collect the random sample $S_\mathcal{A}$ from a nonuniform stationary distribution $\bm{\pi}_\mathcal{A}$. Each answer in $S_\mathcal{A}$  is sampled with a different probability according to its semantic similarity to the query graph, hence different answers should contribute differently to the estimation. Therefore, we cannot apply the estimators designed for a uniform sample, e.g., \cite{Hellerstein1997,Wu2010,Li2016}
\footnote{\scriptsize{\noindent The estimators \cite{Hellerstein1997,Wu2010,Li2016} on relational databases are unbiased only for a uniform sample,
i.e., each tuple is selected with the same probability from a relation.}}.

As an alternative, we provide estimators $\hat{f}_a$ for $\{\sf{SUM}$, $\sf{COUNT}$, $\sf{AVG}\}$ based on the \textit{Horvitz-Thompson estimators} \cite{Horvitz1952} as follows, where $\pi'_i\in \bm{\pi}_\mathcal{A}$ is the visiting probability of each answer $u^t_i\in S_\mathcal{A}$, and $S^+_\mathcal{A}=\{u^t_i\in S_\mathcal{A}:s_i\geq \tau\}$ is the set of answers in $S_\mathcal{A}$ having semantic similarity $\geq \tau$.
\begin{equation}
\small
\label{eq:HT}
\hat{f}_a^{{\rm sum}}(S_\mathcal{A})=\frac{1}{|S^+_\mathcal{A}|}\sum\limits_{u^t_i\in S^+_\mathcal{A}}\frac{u^t_i.a}{\pi'_i}
\end{equation}
\vspace{-0.1cm}
\begin{equation}
\small
\label{eq:HT1}
\hat{f}_a^{{\rm count}}(S_\mathcal{A})=\frac{1}{|S^+_\mathcal{A}|}\sum\limits_{u^t_i\in S^+_\mathcal{A}}\frac{1}{\pi'_i}
\end{equation}
\vspace{-0.1cm}
\begin{equation}
\small
\label{eq:HT2}
\hat{f}_a^{{\rm avg}}(S_\mathcal{A})=\frac{\hat{f}_a^{{\rm sum}}(S_\mathcal{A})}{\hat{f}_a^{{\rm count}}(S_\mathcal{A})}=\frac{\sum_{u^t_i\in S^+_\mathcal{A}}u^t_i.a/\pi'_i}{\sum_{u^t_i\in S^+_\mathcal{A}}{1/\pi'_i}}
\end{equation}
%
%
%
\begin{myLemma}
\label{lemma:3}
The estimator for $\sf{SUM}$ is unbiased. $E[\hat{f}_a^{{\rm sum}}]=\sum_{u^t_i\in \mathcal{A}^+}u^t_i.a$, where $\mathcal{A}^+\subseteq \mathcal{A}$ is the set of correct answers.
\end{myLemma}
\begin{IEEEproof}
\label{pf:7}
Since each answer $u^t_i\in S^+_\mathcal{A}$ is sampled from the identical stationary distribution $\bm{\pi}_\mathcal{A}$ (Theorem \ref{th:iid}), the expectation of each $u^t_i.a/\pi'_i$ is the same. Now, we have the following derivation to show that $\hat{f}_a^{{\rm sum}}$ is an unbiased estimator.
\end{IEEEproof}
\begin{equation}\nonumber
\small
\begin{split}
E[\hat{f}_a^{{\rm sum}}] &=\frac{1}{|S^+_\mathcal{A}|}\sum\limits_{u^t_i\in S^+_\mathcal{A}}E[\frac{u^t_i.a}{\pi'_i}]
             =\frac{1}{|S^+_\mathcal{A}|}(|S^+_\mathcal{A}|\cdot E[\frac{u^t_i.a}{\pi'_i}]) \\
             &=\sum_{u^t_i\in \mathcal{A}^+}\frac{u^t_i.a}{\pi'_i}\cdot \pi'_i
             =\sum_{u^t_i\in \mathcal{A}^+}u^t_i.a
\end{split}
\end{equation}
\begin{myLemma}
\label{lemma:4}
The estimator for $\sf{COUNT}$ is unbiased.
\end{myLemma}
%
\begin{IEEEproof}
\label{pf:8}
Similar to Lemma \ref{lemma:3}, we have the following derivation, so that the estimator for $\sf{COUNT}$ is unbiased.
\end{IEEEproof}
\begin{equation}\nonumber
\small
E[\hat{f}_a^{{\rm count}}] =\frac{1}{|S^+_\mathcal{A}|}\sum\limits_{u^t_i\in S^+_\mathcal{A}}E[\frac{1}{\pi'_i}]
=\frac{1}{|S^+_\mathcal{A}|}\cdot (|S^+_\mathcal{A}|\cdot |\mathcal{A}^+|)=|\mathcal{A}^+|
\end{equation}
\begin{myLemma}
\label{lemma:5}
The estimator for $\sf{AVG}$ is consistent.
\end{myLemma}
\begin{IEEEproof}
\label{pf:9}
We prove that $\hat{f}_a^{{\rm avg}}$ is consistent through two steps by the \textit{Strong Law of Large Numbers} (SLLN) \cite{Jones2004,Lee2012,Li2019}. 

\noindent\textbf{Step 1.} We first transform $\hat{f}_a^{{\rm avg}}$ according to the  \textit{importance sampling framework} \cite{Hansen1943} by setting a weight $w_i=U_i/\pi'_i$ for each $u^t_i\in S^+_\mathcal{A}$, where $U_i=\frac{1}{|S^+_\mathcal{A}|}$ ($U$ is a uniform distribution).
\vspace{-0.15cm}
\begin{equation}\nonumber
\small
\begin{split}
\hat{f}_a^{{\rm avg}}=\frac{\sum_{u^t_i\in S^+_\mathcal{A}}\frac{u^t_i.a}{\pi'_i}}{\sum_{u^t_i\in S^+_\mathcal{A}}\frac{1}{\pi'_i}}=\frac{\frac{1}{|S^+_\mathcal{A}|}\sum_{u^t_i\in S^+_\mathcal{A}} \frac{U_i}{\pi'_i}\cdot u^t_i.a}{\frac{1}{|S^+_\mathcal{A}|}\sum_{u^t_i\in S^+_\mathcal{A}} \frac{U_i}{\pi'_i}}=\frac{\mu(w_i\cdot u^t_i.a)}{\mu(w_i)}\ .
\end{split}
\end{equation}

\noindent\textbf{Step 2.} According to SLLN, the mean ($\mu$) of any function of samples collected from a stationary distribution approximates to the expectation of this function over the stationary distribution \cite{Li2019}, that is, $\mu(g)\rightarrow E_\pi[g]$. So we have the following:
\vspace{-0.25cm}
\begin{equation}\nonumber
\small
\begin{split}
\frac{\mu(w_i\cdot u^t_i.a)}{\mu(w_i)}\rightarrow \frac{E_{\bm{\pi}_\mathcal{A}}[w_i\cdot u^t_i.a]}{E_{\bm{\pi}_\mathcal{A}}[w_i]}=\frac{E_U[u^t_i.a]}{1}=\frac{\sum_{u^t_i\in \mathcal{A}^+} u^t_i.a}{|\mathcal{A}^+|}\ .
\end{split}
\end{equation}
Thus, $\sf{AVG}$'s estimator is consistent, because its expectation converges almost surely to the true mean of all $u^t_i\in \mathcal{A}^+$.
\end{IEEEproof}

\vspace{0.05cm}
\noindent\underline{\textbf{Remarks.}} The estimators in Eq. \ref{eq:HT}-\ref{eq:HT2} are mean-like expressions, so we can apply the \textit{Central Limit Theorem} to provide accuracy guarantee (discussed in \S \ref{accuracy}). It is hard to form a mean-like estimator for extreme functions (e.g., ${\sf MAX}$, ${\sf MIN}$), so we cannot easily provide the CLT-based accuracy guarantee for them. Instead, extreme estimation based on Extreme Value Theory (EVT) \cite{Coles2001} could be an alternative direction. If we can control the sample's distribution to follow Generalized Extreme Value (GEV) or Generalized Pareto distribution (GPD), then we may be able to do the EVT-based estimation. We keep this as an interesting open problem for future work.

%
\subsubsection{Correctness Validation for the Sample}
\label{similarity}
Although our semantic-aware sampling is designed to collect answers with greater semantic similarities as the sample $S_\mathcal{A}$, it is still likely to get a few answers with lower semantic similarity in $S_\mathcal{A}$ due to the randomness of sampling. For the query graph in Figure \ref{fig:AQG}, we find that 12\% answers of $S_\mathcal{A}$ have semantic similarity $<$ $\tau$ on average when $\tau=0.85$. This would affect the accuracy if we directly regard all answers in $S_\mathcal{A}$ as correct and apply Eq. \ref{eq:HT}-\ref{eq:HT2} for estimation. So, we need a method to quickly validate the correctness for answers in $S_\mathcal{A}$ before the estimation. This is exactly what we have done in Eq. \ref{eq:HT}-\ref{eq:HT2}: We first compute $S^+_\mathcal{A}=\{u^t_i\in S_\mathcal{A}:s_i\geq \tau\}$. In experimental study (\S \ref{scale}), we show that our method accounts for 27\% of total query time,
but offers 9X effectiveness improvement on average.

\vspace{0.05cm}
\noindent\underline{\textbf{Basic idea.}} A straightforward way of correctness validation for an answer $u^t_i$ $\in$ $S_\mathcal{A}$ is to find the subgraph match $M(u^t_i)$ = $\overline{u^su^t_i}$ having the greatest semantic similarity to the query graph. Enumerating all subgraph matches, however, is computationally expensive. Instead, we present a heuristic method to find a subgraph match with a higher likelihood of having the greatest semantic similarity and check
its correctness, thus pruning other matches to improve the efficiency. We also discuss the impact on accuracy due to our heuristic method.

\vspace{0.05cm}
\noindent\underline{\textbf{Greedy search heuristic.}} We start search from the mapping node $u^s$ by considering it as the current node. We also initialize a candidate set with $u^s$'s neighbors. Every time, we select the node $u$ from the candidate set that has the highest visiting probability $\pi$, and make it the current node. We update the candidate set by adding $u$'s neighbors that were not current nodes in the past and remove $u$ from it. We continue to search and record all the possible paths from $u^s$ till an answer $u^t_i$ becomes the current node. The found path $\overline{u^su^t_i}$ is used as the (heuristically) best subgraph match for correctness checking.

\vspace{0.05cm}
\noindent\underline{\textbf{Effectiveness analysis.}}  In the context of our greedy search heuristic, a false positive indicates that an incorrect answer $u^t_i$ is reported as a correct answer. False positive would strongly impact the estimation accuracy.
Fortunately, we can guarantee that no false positive would happen in our greedy search. Consider an incorrect answer $u^t_i$, all subgraph matches $M(u^t_i)$ of $u^t_i$ must have a semantic similarity $<\tau$ by definition. Hence, we never include this $u^t_i$ in $S^+_\mathcal{A}$, no matter what subgraph match of $u^t_i$ is returned via our greedy search. On the other hand, a false negative indicates that a correct answer $u^t_i$ is reported as an incorrect answer. This can happen when the subgraph match found so far is not the optimal one. To reduce the chances of false negative, we introduce a \textit{repeat factor} $r$ in our correctness validation. Specifically, our greedy search continues searching until $r$ paths from $u^s$ to $u^t_i$ are found and returns the correctness of the one with the greatest semantic similarity. The larger $r$ is,
the lower probability of false negative happens, but this also consumes more time for correctness validation. In \S \ref{sensitivity}, we show that a balance between efficiency and effectiveness is achieved when $r=3$.


\subsection{Accuracy Guarantee}
\label{accuracy}
After correctness validation, we obtain the approximate result  $\hat{V}$ by Eq. \ref{eq:HT}-\ref{eq:HT2}. Given a user-specific error bound $e_b$, we guarantee that the relative error $|\hat{V}-V|/V\leq e_b$, where $V$ is the ground truth. To achieve this, we compute a confidence interval CI $=\hat{V}\pm\varepsilon$ (at a user-specific confidence level $1-\alpha$) to quantify the quality of $\hat{V}$. This CI states that $V$ is covered by an estimated range $\hat{V}\pm\varepsilon$ with probability $1-\alpha$. The half width $\varepsilon$ is called the Margin of Error (MoE). We first prove that $|\hat{V}-V|/V\leq e_b$ holds with probability $1-\alpha$ if $\varepsilon\leq \hat{V}\cdot e_b/(1+e_b)$. Then, we show how to compute $\varepsilon$ efficiently.

\vspace{0.05cm}
\begin{myTheorem}
\label{th:accuracy}
If the MoE $\varepsilon$ of the CI satisfies $\varepsilon\leq \frac{\hat{V}\cdot e_b}{1+e_b}$, then the relative error of the approximate result is upper bounded by $e_b$, denoted as $|\hat{V}-V|/V\leq e_b$, with probability $1-\alpha$.
\end{myTheorem}
\vspace{0.05cm}
\begin{IEEEproof}
\label{pf:accuracy}
Given the CI $=[\hat{V}-\varepsilon,\hat{V}+\varepsilon]$ at the confidence level $1-\alpha$, we prove this theorem in two steps.

\vspace{0.05cm}
\noindent\textbf{Step 1.} Suppose that $V$ is located in the CI’s right half-width, i.e., $\hat{V}$ $\leq$ $V$ $\leq$ $\hat{V}+\varepsilon$, then we have the following derivation and $(V-\hat{V})/V$ $\leq$ $e_b$ holds if $\varepsilon/\hat{V}$ $\leq$ $e_b$ (i.e., $\varepsilon$ $\leq$ $\hat{V}\cdot e_b$).
\begin{equation}\nonumber
\small
(V-\hat{V})/V\leq (V-\hat{V})/\hat{V}\leq \varepsilon/\hat{V}
\vspace{-2mm}
\end{equation}

\noindent\textbf{Step 2.} Suppose that $V$ is located in CI's left half-width, i.e., $\hat{V}-\varepsilon\leq V\leq \hat{V}$. Then we say that $(\hat{V}-V)/V\leq e_b$ holds if $\varepsilon/(\hat{V}-\varepsilon)\leq e_b$ (i.e., $\varepsilon\leq \frac{\hat{V}\cdot e_b}{1+e_b}$) by the following derivation.
\begin{equation}\nonumber
\small
(\hat{V}-V)/V\leq \varepsilon/V\leq \varepsilon/(\hat{V}-\varepsilon)
\end{equation}

In summary, $\frac{|\hat{V}-V|}{V}\leq e_b$ holds if $\varepsilon\leq \frac{\hat{V}\cdot e_b}{1+e_b}$, because $\frac{\hat{V}\cdot e_b}{1+e_b}\leq \hat{V}\cdot e_b$ is a tighter bound. Finally, $\hat{V}$ is in CI with probability $1-\alpha$,
indicating that the above holds with probability $1-\alpha$.
\end{IEEEproof}

\vspace{0.05cm}
\noindent\underline{\textbf{Confidence interval calculation.}} The \textit{Central Limit Theorem} (CLT) is applied to compute the confidence interval \cite{Wu2010}. We take ${\sf SUM}$ as an example to show the basic idea. Consider $V_i= u^t_i.a/\pi'_i$ as a random variable w.r.t. each answer $u^t_i\in S^+_\mathcal{A}$. The approximate result $\hat{V}$ can be viewed as the mean of a set of random variables: $\{V_i|u^t_i\in S^+_\mathcal{A}\}$ according to Eq. \ref{eq:HT}. From CLT, we know that if a point estimator takes the form of the mean of i.i.d. random variables, then it follows a normal distribution \cite{Gao2019}. Hence, we have $\hat{V}\sim N(\mu_{\hat{V}},\sigma^2_{\hat{V}})$, and the MoE $\varepsilon$ of the confidence interval $\hat{V}\pm \varepsilon$ at a confidence level $1-\alpha$ can be calculated based on CLT as follows.
\begin{equation}
\small
\label{eq:moe}
\varepsilon=z_{\alpha/2}\cdot \sigma_{\hat{V}}
\end{equation}

Here, $z_{\alpha/2}$ is the normal critical value with right-tail probability $\alpha/2$, which can be obtained from a standard normal table.
We use the \textit{Bag of Little Bootstrap} (BLB) \cite{Kleiner2014} to estimate 
$\sigma_{\hat{V}}$.

\vspace{0.05cm}
\noindent\textbf{\textit{Bag of little bootstrap}.}
We initialize the desired sample size $N=\lambda\cdot |\mathcal{A}|$ as a fraction of the candidate answer set $\mathcal{A}$, where $\lambda$ is the desired sample ratio.
(1) BLB collects $t$ small samples $\{S_i|i=1,\cdots,t\}$ from $\mathcal{A}$ to form the random sample $S_\mathcal{A}=\bigcup_{i=1}^t S_i$, each $S_i$ has a size $|S_i|=N^m$, so we have $|S_\mathcal{A}|=t\cdot N^m$. The scale factor $m\in [0.5,1]$ is used in \cite{Kleiner2014} to satisfy $t\cdot N^m<N$. (2) For each small sample $S_i$, BLB estimates $\sigma_{\hat{V}}$ by a standard bootstrap \cite{Shao2012} (given below) and compute an MoE $\varepsilon_i$ by Eq. \ref{eq:moe}. (3) Given a set of MoE $\{\varepsilon_1\cdots \varepsilon_t\}$ for $t$ small samples, BLB computes the final $\varepsilon=\sum \varepsilon_i/t$.

\vspace{0.05cm}
\noindent\textbf{\textit{Bootstrap}.}
(1) We collect $B$ resamples from $S_\mathcal{A}$ with replacement, each resample contains $|S_\mathcal{A}|$ answers.
(2) We compute the approximate result for each resample as $\{\hat{V}^*_1...\hat{V}^*_B\}$. (3) Bootstrap takes the empirical distribution of $\{\hat{V}^*_1...\hat{V}^*_B\}$ as an approximation to $N(\mu_{\hat{V}},\sigma^2_{\hat{V}})$, so we estimate $\sigma_{\hat{V}}$ by Eq. \ref{eq:mu}. 
\begin{equation}
\small
\label{eq:mu}
\mu_{\hat{V}}=\sum \hat{V}^*_i/B\ ,\quad \sigma^2_{\hat{V}}=\sum (\hat{V}^*_i-\mu_{\hat{V}})^2/(B-1)
\end{equation}

We find that $\lambda=0.3$ is enough to obtain a good result (\S \ref{sensitivity}) for $t$ $\geq$ $3$, $m$=$0.6$, and $B$ $\geq$ $50$ as suggested in \cite{Kleiner2014}.

\vspace{0.05cm}
\noindent\underline{\textbf{Configuration of $|\Delta S_{\mathcal{A}}|$.}} We terminate the query when the MoE $\varepsilon\leq \frac{\hat{V}\cdot e_b}{1+e_b}$ (Theorem \ref{th:accuracy}). Otherwise, we update $S_\mathcal{A}$ by additional answers, i.e., $S_\mathcal{A}=S_\mathcal{A}\cup\Delta S_\mathcal{A}$, and continue the estimation until we obtain a small enough $\varepsilon$. Intuitively, we need a large $|\Delta S_\mathcal{A}|$ when $\varepsilon$ is large. Otherwise, a small $|\Delta S_\mathcal{A}|$ would be sufficient. To achieve the right balance, we present an error-based method that automatically configures $|\Delta S_\mathcal{A}|$.

Consider an MoE $\varepsilon>\frac{\hat{V}\cdot e_b}{1+e_b}$. We use $\varepsilon/\frac{\hat{V}\cdot e_b}{1+e_b}$ to denote how far $\varepsilon$ is away from the desired value $\frac{\hat{V}\cdot e_b}{1+e_b}$. The larger $\varepsilon/\frac{\hat{V}\cdot e_b}{1+e_b}$ is, the more answers that $\Delta S_\mathcal{A}$ requires. Ideally, if we can reduce $\varepsilon$ to a new $\varepsilon'$ by at least $\varepsilon/\frac{\hat{V}\cdot e_b}{1+e_b}$ times, we can satisfy $\varepsilon'\leq \frac{\hat{V}\cdot e_b}{1+e_b}$. According to Eq. \ref{eq:moe}, reducing $\varepsilon$ by $\varepsilon/\frac{\hat{V}\cdot e_b}{1+e_b}$ times is equivalent to reducing $\sigma_{\hat{V}}$ by $\varepsilon/\frac{\hat{V}\cdot e_b}{1+e_b}$ times. Since $\sigma_{\hat{V}}=\frac{\sigma_V}{\sqrt{N}}$ according to CLT, where $\sigma_V$ is the standard deviation of the population, we say that reducing $\sigma_{\hat{V}}$ by $\varepsilon/\frac{\hat{V}\cdot e_b}{1+e_b}$ times is equivalent to increasing $N$ by $(\varepsilon/\frac{\hat{V}\cdot e_b}{1+e_b})^2$ times. In summary, we can increase $N$ by $(\varepsilon/\frac{\hat{V}\cdot e_b}{1+e_b})^2$ times to reduce $\varepsilon$ by $\varepsilon/\frac{\hat{V}\cdot e_b}{1+e_b}$ times. Hence, we derive $|\Delta S_\mathcal{A}|$ as follows.
\vspace{-0.2cm}
\begin{equation}
\small
\label{eq:delta}
\begin{aligned}
|\Delta S_\mathcal{A}|&= t\cdot [N\cdot (\varepsilon/\frac{\hat{V}\cdot e_b}{1+e_b})^2]^m-t\cdot N^m\\
&=|S_\mathcal{A}|\cdot [(\varepsilon/\frac{\hat{V}\cdot e_b}{1+e_b})^{2m}-1]
\end{aligned}
\end{equation}
\vspace{-0.1cm}

\vspace{-0.1cm}
\begin{myExample}
\label{exp:deltaS}
Let us consider the query in Figure \ref{fig:AQG} (ground truth is $596$). Suppose that we already have a ${\rm CI}=\hat{V}\pm \varepsilon$ with $\hat{V}$ = $578$ and $\varepsilon$ = $6.5$, by a sample with size $|S_\mathcal{A}|$ = $100$. If we set the scale factor $m$ = $0.6$ and the error bound $e_b$ = $0.01$, then we only need to collect $|\Delta S_\mathcal{A}|$ = $100\cdot ((6.5/\frac{578\cdot 0.01}{1.01})^{2\cdot 0.6}-1)\approx 16$ additional answers, which is much smaller than $|S_\mathcal{A}|$.
\end{myExample}

\vspace{0.05cm}
\noindent\underline{\textbf{Interactive refinement of $e_b$.}} In an interactive scenario, a user varies the error bound $e_b$ in runtime, that is, she gradually reduces $e_b$ to achieve more accurate results. Our approach can quickly obtain a new approximate result with a small additional overhead. This is because our error-based sample size configuration method can sense the variation of $e_b$ and update $S_\mathcal{A}$ appropriately via Eq. \ref{eq:delta}. This can be very beneficial to reduce overhead caused by oversampling.

\vspace{0.05cm}
\noindent\underline{\textbf{Remarks}}. The empirical results (\S \ref{sensitivity}) show that the desired sample ratio $\lambda=0.3$ is enough to achieve a good result for $t$ $\geq$ $3$, $m$=$0.6$, and $B$ $\geq$ $50$ as recommended in \cite{Kleiner2014}. 
\subsection{Putting All Together}
\label{alg_blk}
We present the entire algorithm in Algorithm \ref{alg:whole_alg}, which includes an offline KG embedding phase
to generate a predicate vector space $\bm{E}$ (Line 1) and an online ``sampling-estimation'' phase to obtain an approximate aggregate result with an accuracy guarantee (Lines 2-14). In the online phase, given an aggregate query $AQ_G$ = $(Q,f_a)$, we do the following (Lines 2-3): (1) Conduct a semantic-aware random walk on $G$ (within $n$-hops) to reach a stationary distribution $\bm{\pi}$ by using $\bm{E}$, and (2) collect a random sample $S_\mathcal{A}$ from $G$ according to $\bm{\pi}$. Next, we validate the correctness of each answer in $S_\mathcal{A}$ to obtain correct answers $S_\mathcal{A}^+\subseteq S_\mathcal{A}$ (Line 5) and estimate the approximate result $\hat{V}$ by applying the proposed unbiased estimators or consistent estimator on $S_\mathcal{A}^+$ (Line 6). After that, we compute the MoE $\varepsilon$ of a confidence interval CI $=\hat{V}\pm \varepsilon$ at the confidence level $1-\alpha$ (Line 7). Finally, we check the termination condition proved in Theorem \ref{th:accuracy} and return $\hat{V}\pm \varepsilon$ when $\varepsilon\leq \frac{\hat{V}\cdot e_b}{1+e_b}$ (Lines 8-9 \& 14). Otherwise, we enlarge $S_\mathcal{A}$ with an appropriately configured sample size $|\Delta S_\mathcal{A}|$, and repeat above steps to continuously refine $\hat{V}\pm\varepsilon$ (Lines 11-13).

\vspace{0.1cm}
\noindent\textbf{\underline{Remarks}.} The total time of our method consists of the time for sampling ($T_s$) and that of aggregate estimation ($T_e$). We have $T_s=O(|E_{G'}|+N_{ws}+|S_\mathcal{A}|)$, where $|{E_{G'}}|$ is the average number of edges in the scope of our random walk and we compute the transition probability for each such edge, $N_{ws}$ is the average \underline{w}alk \underline{s}teps required for random walk convergence ($N_{ws}\leq 500$ in practice), and $|S_\mathcal{A}|$ is the initial sample size. Next, we have $T_e=O(N_e\cdot(|S_\mathcal{A}|+|S^+_{\mathcal{A}}|+|\Delta S_\mathcal{A}|))$, where $N_e$ is the average number of iterations till the termination condition proved in Theorem \ref{th:accuracy} is satisfied ($N_e\leq 10$ in practice). In each iteration, we validate the correctness of $|S_\mathcal{A}|$ answers (Line 5), estimate based on $|S^+_{\mathcal{A}}|$ correct answers (Lines 6-7), and include $|\Delta S_\mathcal{A}|$ additional answers to refine the approximate result (Lines 11-13).

\begin{algorithm}[t]
\footnotesize
\caption{{\small Approximate Aggregate Query on KGs}}
\label{alg:whole_alg}
\KwData{A KG $G$, an aggregate query $AQ_G=(Q,f_a)$, a user-input error bound $e_b$ and a confidence level $1-\alpha$}
\KwResult{approximate aggregate result with a CI $=\hat{V}\pm \varepsilon$}
$\bm{E}=$ KG\_Embedding\_Model($G$)\;
\tcp{\small{\S \ref{sampling}$\sim$\ref{accuracy}: ``sampling-estimation''}}
$\bm{\pi}=$ randomWalk($G$, $\bm{E}$)\tcc*{\footnotesize{\S \ref{srws}(2)}}
$S_{\mathcal{A}}=$ collectSample($G$, $\bm{\pi}$, $Q.{q^t}$)\tcc*{\footnotesize{\S \ref{srws}(3)}}
\While{${\sf true}$}{
	$S^+_\mathcal{A}=$ correctnessValidate($G$, $S_\mathcal{A}$, $\bm{\pi}$, $\tau$)\tcc*{\footnotesize{\S \ref{similarity}}}
	$\hat{V}=$ estimate($S^+_\mathcal{A}$)\tcc*{\footnotesize{\S \ref{estimator}}}
	$\varepsilon=$ getMoE($S^+_\mathcal{A}$, $1-\alpha$)\tcc*{\footnotesize{\S \ref{accuracy}}}
	\If{$\varepsilon\leq \hat{V}\cdot e_b/(1+e_b)$}{
		\textbf{break}\tcc*{\footnotesize{\S  \ref{accuracy}: Theorem \ref{th:accuracy}}}
	}\Else{
		$|\Delta S_\mathcal{A}|=$ configSampleSize($\varepsilon$,$\hat{V}$, $e_b$)\tcc*{\footnotesize{\S  \ref{accuracy}}}
		$\Delta S_{\mathcal{A}}=$ collectSample($G$, $\bm{\pi}$, $Q.{q^t}$)\;
		$S_\mathcal{A}=S_\mathcal{A}\cup\Delta S_\mathcal{A}$\;
	}
}
\Return CI $=\hat{V}\pm \varepsilon$\;
\end{algorithm}
%

\section{Extensions}
\label{general}
We extend our solution to support more complex queries with filters, GROUP-BY, and different graph shapes.

\vspace{-0.15cm}
\subsection{Aggregate Queries with Filters and GROUP-BY}
\label{range}
\noindent\underline{\textbf{Queries with filters.}} We define queries with filters, e.g., \textit{range queries}, which form a common query type, as follows.

\begin{myDef}
\label{def:filter_query}
\textbf{$AQ_G$ with filters}. An aggregate query with filters is defined as $AQ_G^F=(Q,\bigcup_{i=1}^{n}L_i\leq b_i\leq U_i,f_a)$, where (1) $\bigcup_{i=1}^{n}L_i\leq b_i\leq U_i$ is a set of filters, and (2) for each filter, the attribute $b_i$ of each answer to $Q$ must have a value between a lower ($L_i$) and an upper ($U_i$) bound.
\end{myDef}

\begin{myExample}
\label{exp:filter_query}
Given a query: ``\textit{Find the average price of cars produced in Germany with a fuel\_economy between 25 and 30 MPG}'', we can form the $AQ_G^F$ of this query by adding a filter 25$\leq$ ${\sf fuel\_economy}$ $\leq$30 on the aggregate query in Figure \ref{fig:AQG}.
\end{myExample}
%
We support $AQ_G^F$ by adding a filter operation in the correctness validation (\S \ref{similarity}): only the answer $u^t_i$ satisfying the filter condition and its semantic similarity $s_i\geq \tau$ is a correct answer, i.e., $c(u^t_i)=(L\leq u^t_i.b\leq U$ $\&\&$ $s_i\geq \tau)?1:0$.

\vspace{0.05cm}
\noindent\underline{\textbf{Queries with GROUP-BY.}} We support queries with GROUP-BY to return answers in groups.
For example, 
``\textit{How many Spanish soccer players of each age group are there?}".
In particular, when GROUP-BY is applied on the query node that serves as a target node,
we only need to divide the collected sample of soccer players into different groups according to their ages,
then we estimate and return the approximate result for each group.

%

\begin{figure}
\vspace{-0.3cm}
\setlength{\abovecaptionskip}{0.1cm}
\centerline{\includegraphics[scale=0.67]{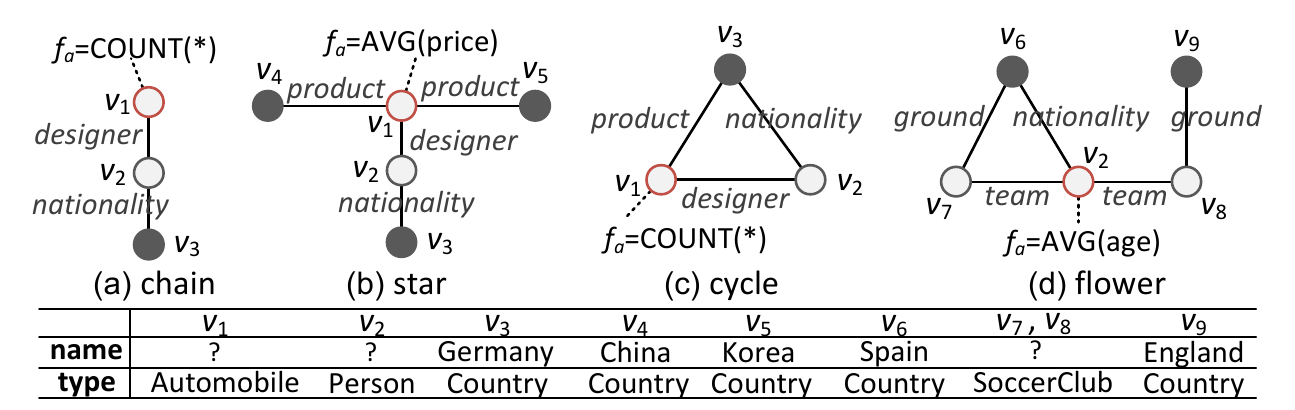}}
\caption{\small Complex queries with different shapes}
\label{fig:complex_queries}
\end{figure}
%

\subsection{Aggregate Queries with Different Shapes}
\label{shapes}
Users generally form different shaped query graphs, 
e.g., the commonly used chain, star, cycle, and flower shapes \cite{Bonifati2017}.
We first extend our solution to chain-shaped queries, then we introduce a general method to support other shapes.

\vspace{0.05cm}
\noindent\underline{\textbf{Chain-shaped queries.}}
A chain-shaped query is defined as $AQ_G^C=(Q^C,f_a)$: (1) $Q^C$ is a multi-hop path from a specific node $q^s$ to a target node $q^t$. (2) For $q^s$, both types and name are known. All other nodes after $q^s$ (including $q^t$) are unknown nodes (only their node types are given). (3) the simple query is a special case of $AQ_G^C$ with an edge between $q^s$ and $q^t$.
%

Figure \ref{fig:complex_queries} (a) shows a chain-shaped query: ``\textit{How many cars are designed by German designers?}".
With this as an example, (1) we start a semantic-aware sampling from the specific entity ${\sf Germany}$ to get a
temporary sample of ${\sf Person}$ entities. (2) For each person, we do the second semantic-aware sampling to collect the final sample
of ${\sf Automobile}$ entities till enough automobiles are obtained (each second sampling is run as a thread).
(3) Suppose that each ${\sf Person}$ entity $u_i$ in ${\sf Germany}$'s $n$-bounded subgraph has a stationary visiting probability $\pi'_i$ and each ${\sf Automobile}$ entity $u_j$ belonging to $u_i$'s $n$-bounded subgraph has a stationary visiting probability $\pi'_j$, then this automobile $u_j$ can be sampled with probability $\pi'=\pi'_i\cdot \pi'_j$, where $\sum\pi'=\sum_{i,j}\pi'_i\cdot \pi'_j=\sum_i(\pi'_i\cdot\sum_j\pi'_j)=1$. If $u_j$ appears multiple times in the final sample, all
such $\pi'$ corresponding to $u_j$ are accumulated to derive its final sampling probability. (4) We estimate the approximate results based on the aforementioned final sample following \S \ref{accuracy}.

\vspace{0.05cm}
\noindent\underline{\textbf{Queries with other shapes.}} Given the simple and chain-shaped queries, we can form more complex queries with other shapes (Figure \ref{fig:complex_queries} (b-d)). For example, the star-shaped query aims to ``\textit{find the average price of cars produced in both China and Korea, and designed by German
designers}'', which consists of two simple queries and a chain-shaped query. We adopt the ``\textit{decomposition-assembly}" framework \cite{Yang2016,Wang2020} to support these queries: (1) We decompose a complex query into a set of simple or chain-shaped queries that share the same target entity, denoted by $\{Q_1,\cdots,Q_n\}$, (2) we collect a random sample $S_\mathcal{A}^i$ for each $Q_i$, (3) we use the intersection $S_\mathcal{A}=\bigcap_{i=1}^{n}S^i_\mathcal{A}$ as the random sample for the original complex query, and (4) we iteratively estimate the approximate result based on $S_\mathcal{A}$ until we obtain an accurate enough result.

\begin{table}
\setlength{\abovecaptionskip}{0.1cm}
\setstretch{1}
\fontsize{6.3pt}{2.6mm}\selectfont
  \centering
  \tabcaption{Statistics of our datasets}
  \begin{tabular} {c||c|c|c}
   Datasets & DBpedia & Freebase & YAGO2\\
   \hline \hline
   \#Nodes & 4,521,912 & 5,706,539 & 7,308,072\\
   \hline
   \# Edges & 15,045,801 & 48,724,743 & 36,624,106\\
   \hline
   \# Node-Types & 359 & 11,666 & 6,543\\
   \hline
   \# Edge-Predicates & 676 & 5118 & 101  \\
  \end{tabular}
  \label{tab:dataset}
\end{table}

\begin{table*}
\vspace{-0.6cm}
\setstretch{1}
\fontsize{6.2pt}{2.6mm}\selectfont
\setlength{\abovecaptionskip}{0.1cm}
    \centering
    \tabcaption{\small Ten query examples out of all queries (10/400) that we used in \S \ref{experiment}}
    \begin{tabular}{c||c|c|c}
	\textbf{QID} & \textbf{Queries} & \textbf{Category \& selectivity (\%)}	 & \textbf{Aggregate function }\bm{$f_a$}\\
	\hline \hline
	Q1 & \textit{How many cars are produced in Germany?} & Simple, 3.01\% & COUNT(*) \\
    \hline
    Q2 & \textit{What's the average price of cars that are produced in Germany?} & Simple, 3.01\% & AVG(price) \\
    \hline
    Q3 & \textit{What's the average price of cars produced in Germany with a fuel\_economy between 25MPG and 30MPG?} & Simple+Filter, 0.51\%  & AVG(price) \\
    \hline
    Q4 & \textit{How many football clubs are there in each country?} & Simple+GROUP-BY & COUNT(*) \\
    \hline
    Q5 & \textit{How many languages are spoken in Nigeria?} & Simple, 69.59\% & COUNT(*) \\
    \hline
    Q6 & \textit{What's the total box office of the movies that were directed by Steven Spielberg?} & Simple, 0.05\% & SUM(box\_office) \\
    \hline
    Q7 & \textit{How many museums are there in England?} & Simple, 3.23\% & COUNT(*) \\
    \hline
    Q8 & \textit{What's the average population of China's cities?} & Simple, 2.91\% & AVG(population) \\
    \hline
    Q9 & \textit{How many soccer players were born in Spain and played for Barcelona FC?} & Star, 0.12\% & COUNT(*) \\
    \hline
    Q10 & \textit{How many cars are designed by German designers?} & Chain, 2.91\% & COUNT(*) \\
    \end{tabular}
    \label{tab:query}
    \vspace{-0.5cm}
\end{table*}

\vspace{-0.1cm}
\section{Related Work}
\label{previous_work}
\noindent\underline{\textbf{Online aggregation.}} Online aggregation (OLA) was first proposed in \cite{Hellerstein1997}, which is a sampling-based technology to return approximate aggregate results on relational data. Much follow-up work has continued over the years, including (1) OLA over joins, group-by \cite{Haas1999,Jermaine2007,Luo2002,Li2016,ParkMSW18,AcharyaGP00,DingHCC016}, (2) OLA for distributed environments \cite{Shi2012,Wu2009,Condie2010,Condie2010a,Pansare2011,Zeng2015}, and (3) OLA with multi queries optimization \cite{Wu2010,Wang2014a}. None of above work can be deployed directly to answer aggregate queries on knowledge graphs, because a knowledge graph’s \textit{schema-flexible} nature is different from relational data. Hence, \textit{we design a new semantic-aware sampling suitable for knowledge graphs, and unbiased (or consistent) estimators for the non-uniform sample}.

\vspace{0.05cm}
\noindent\underline{\textbf{Graph query on knowledge graphs.}} Graph matching is widely used for querying KGs \cite{ZhengLZHZ16,Zou2011,Cheng2008,Fan2010a,Ma2014,Khan2013,Khan2011,Jin2015,Zheng2013,Yang2014,Yang2016,Zheng2016, Wang2020}. They do not consider aggregate queries, except \cite{Zou2011,HuDD20,UngerBLNGC12,HoffnerLU16,HuDYY18}. Since \cite{Zou2011,HuDD20,UngerBLNGC12,HoffnerLU16,HuDYY18} answer aggregate queries based on the factoid queries (often via SPARQL aggregate queries), they suffer from the issues mentioned in \S \ref{intro}. \textit{We propose a novel ``sampling-estimation'' model to answer aggregate queries more  effectively and efficiently}.

\vspace{0.05cm}
\noindent\underline{\textbf{Aggregate query on knowledge graphs.}} Recently, \cite{Li2020} considers aggregate queries on knowledge graphs: It collects candidate entities via link prediction, and computes the aggregate result based on them. Unlike ours, \textit{it does not support the edge-to-path mapping based on semantic similarity, and thus reports lower-quality answers} 
(\S \ref{experiment}). In \cite{Li2020}, the user cannot specify the relative error bound and confidence level, and can neither update them interactively. Finally, unlike our extensions, \cite{Li2020} performs aggregation only for simple queries. 
\vspace{-0.1cm}
\section{Experimental Results}
\label{experiment}
We evaluate (1) effectiveness and efficiency for simple and complex queries with filters, GROUP-BY, and different shapes (\S \ref{effective}), (2) effect of  each step (\S \ref{scale}), and (3) interactive performance and parameter sensitivity (\S \ref{sensitivity}). Our code \cite{code} were implemented in Java1.8 and run on a 2.1GHZ, 64GB memory AMD-6272 server (CentOS Linux).
\vspace{-0.1cm}
\subsection{Environment Setup}
\label{setup}
\noindent\underline{\textbf{Datasets.}} We used three real-world datasets as shown in Table \ref{tab:dataset}. (1) \textbf{\textit{DBpedia}} \cite{Lehmann2015} is an open-domain KG, which is constructed from Wikipedia. (2) \textbf{\textit{Freebase}} \cite{Bollacker2008} is a KG harvested from many sources, including individual, user-submitted wiki contributions. Since we assume that each entity has a name, we used a Freebase-Wikipedia mapping file \cite{freebaselinks} to filter 5.7M entities, each entity has a name from Wikipedia. (3) \textbf{\textit{YAGO2}} \cite{Hoffart2013} is a KG with information from the Wikipedia, WordNet, and GeoNames. We used the ${\sf CORE}$ portion of YAGO (excluding GeoName) as our dataset. Since we focused on aggregate queries, we complemented the numerical attributes of entities via web crawling. For example, we added attributes, e.g., ${\sf price}$, ${\sf horsepower}$, etc. (from ${\sf edmunds.com}$) to automobiles, added ${\sf age}$, ${\sf transfer\_value}$, etc. (from ${\sf sofifa.com}$) to soccer players, and added ${\sf ratings}$, ${\sf box\_office}$, etc. (from ${\sf IMDB}$) to movies.

\vspace{0.05cm}
\noindent\underline{\textbf{Query workload.}} We tested both {\em benchmark} and {\em synthetic queries}. Table \ref{tab:query} shows 10 examples (out of 400 queries) in our experiments. (1) \textbf{\textit{QALD-4}} \cite{qald} is a benchmark for factoid queries on \textit{DBpedia}. We selected factoid queries from QALD-4 as seeds to form ${\sf COUNT}$, ${\sf AVG}$, and ${\sf SUM}$ queries through a simple modification (e.g., Q1-Q2). For instance, we changed the query ``\textit{Find all cars produced in Germany}'' to ``\textit{How many cars are produced in Germany?}'' (Q1). We also added filters and GROUP-BY to these queries to form more queries (e.g., Q3-Q4).  (2) \textbf{\textit{WebQuestions}} \cite{Berant2013} is a benchmark for \textit{Freebase}. We applied the same method on WebQuestions to form ${\sf COUNT}$, ${\sf AVG}$, and ${\sf SUM}$ queries, as well as queries with filters and GROUP-BY (e.g., Q5-Q6). (3) \textbf{\textit{Synthetic queries}} were generated to evaluate our approach on {\em YAGO2} (e.g., Q7-Q8). Moreover, we generated complex queries with different shapes (e.g., Q9-Q10). According to \cite{Bonifati2017}, most of the real-world queries have small number of edges, e.g., 73.8\% and 81.2\% of queries for DBpedia have the number of edges $\leq$ 2 and 4, respectively. Therefore, we formed 2-hop chain-shaped queries (e.g., Q10) and used them together with simple queries to generate queries with other shapes (e.g., Fig. \ref{fig:complex_queries} (b-c)). In Table \ref{tab:query}, we show the selectivity of each query, defined as the percentage of correct answers, over all candidate answers. The average selectivity of all queries is 6.39\%, whereas the largest selectivity of any of our queries is around 70\%. {\em Thus, our approach can support both high and low-selectivity queries.}
%

\vspace{0.05cm}
\noindent\underline{\textbf{Metrics.}} We run each query for 5 times, and used \textit{average relative error} and \textit{average response time} as the metrics for effectiveness and efficiency, respectively. Initially, 38.5\% (154/400) of our queries, which were formed from seed queries of \textit{QALD-4} and \textit{WebQuestions} (e.g., Q1-Q3, Q5-Q6 in Table \ref{tab:query}), already had human-annotated ground truth. For the remaining 246 queries, we conducted crowdsourcing with the \textit{Baidu Data Crowdsoursing Platform} (\url{https://zhongbao.baidu.com}) to obtain human-annotated ground truth. Specifically, for each query graph $Q$ of an aggregate query $AQ_G$, we extracted all the possible schemas between a target entity $q^t\in Q$ and a specific entity $q^s\in Q$ from the KG $G$ and send them to 10 annotators; we asked them to annotate all those schemas which are semantically similar to the given predicate $L_Q(e)$ of edge $e=q^sq^t$. Next, we collected annotations from 10 annotators and used the intersection of them as the candidate schemas. Finally, we extracted all the entities with the same type as $q^t$ that match with these candidate schemas as the human-annotated correct answers to $Q$, and applied the aggregate function on them to obtain the human-annotated ground truth to $AQ_G$. {\em In summary, we have human-annotated ground truth for all our queries.
In our experiments, we measured the effectiveness in two ways for all our queries: based on human-annotated ground truth (HA-GT) and semantic similarity-based ground truth (i.e., $\tau$-relevant ground truth ($\tau$-GT))}.

\begin{table}
\setlength{\abovecaptionskip}{0.1cm}
\setstretch{1}
\fontsize{6.2pt}{2.6mm}\selectfont
  \centering
  \tabcaption{\small Average Jaccard similarity (AJS) between the human-annotated and $\tau$-relevant correct answers and its variance (Var)}
  \begin{tabular} {c||c|c|c|c|c|c|c|c}
   \textbf{Threshold} $\tau$ & 0.6 & 0.65 & 0.7 & 0.75 & 0.8 & 0.85 & 0.9 & 0.95 \\
   \hline \hline
   DBpedia-AJS & 0.64 & 0.71 & 0.74 & 0.78 & 0.88 & \textbf{0.95} & 0.83 & 0.64\\
   DBpedia-Var & 0.084 & 0.080 & 0.079 & 0.043 & 0.023 & \textbf{0.011} & 0.121 & 0.184\\
   \hline
   Freebase-AJS & 0.68 & 0.74 & 0.80 & 0.85 & \textbf{0.96} & 0.91 & 0.82 & 0.70\\
   Freebase-Var & 0.227 & 0.215 & 0.208 & 0.117 & 0.061 & \textbf{0.029} & 0.308 & 0.502\\
   \hline
   Yago2-AJS & 0.52 & 0.63 & 0.70 & 0.82 & \textbf{0.93} & 0.88 & 0.77 & 0.59\\
   Yago2-Var & 0.042 & 0.046 & 0.045 & 0.025 & \textbf{0.013} & 0.015 & 0.66 & 0.108\\
  \end{tabular}
  \label{tab:jaccard}
\end{table}

\begin{table*}
\vspace{-0.6cm}
\setlength{\abovecaptionskip}{0.05cm}
\setstretch{1}
\fontsize{6.2pt}{2.6mm}\selectfont
    \centering
    \caption{\small Effectiveness: relative error (\%) for different query shapes over all datasets and all queries ($\tau$-relevant ground truth)}
\begin{tabular}{c||c|c|c|c|c||c|c|c|c|c||c|c|c|c|c}
\multirow{2}{*}{\textbf{Method}} & \multicolumn{5}{c||}{\textbf{DBpedia}} & \multicolumn{5}{c||}{\textbf{Freebase}} & \multicolumn{5}{c}{\textbf{Yago2}} \\
 & Simple & Chain & Star & Cycle & Flower & Simple & Chain & Star & Cycle & Flower & Simple & Chain & Star & Cycle & Flower  \\ \hline \hline
\textbf{Ours}   &    \textbf{0.84 }      &   \textbf{0.33}     &   \textbf{1.65}     &    \textbf{0.72}      &   \textbf{0.95}    &    \textbf{0.86}    &  \textbf{0.62}    &  \textbf{0.91}    &  \textbf{0.71}     &     \textbf{0.98}     &    \textbf{0.54}    &    \textbf{0.85}    &   \textbf{0.30}       &   \textbf{0.75}      &   \textbf{0.92}      \\ \hline
EAQ    &    20.02       &    -    &    -    &    -    &    -    &   17.74     &    -      &     -   &   -    &     -     &  14.72    &    -    &    - &      -  &    -     \\ \hline
GraB  &     8.08       &   29.10     &    18.69    &    19.24    &    16.38    &   12.84     &     18.47     &     18.91   &    23.13    &    18.87    &   8.57    &       19.23 &   17.53  &   24.61     &   12.87    \\ \hline
QGA  &    17.94     &    31.37    &    38.49    &     24.17   &   31.01     &   34.63    &     29.68     &     34.12   &   21.76     &   20.51     &   19.46    &       22.38 &   32.21  &     42.19   &  32.54     \\ \hline
SGQ   &     10.67     &   12.08     &    16.13    &   9.98     &    12.11    &   6.96     &     5.54     &     14.74   &   13.15     &    15.61    &   7.97    &    14.03    &   10.01  &   15.77     &   8.52    \\ \hline
JENA &    16.60     &    29.37    &    42.03    &    23.09    &    21.32    &   28.17     &     29.68     &     34.77   &    22.76    &    21.24    &   11.30    &   26.38     &   27.38  &   41.03     &    39.47   \\ \hline
Virtuoso &  16.60     &   29.37     &    42.03    &    23.09    &     21.32   &   28.17     &    29.68      &     34.77   &    22.76    &    21.24    &   11.30    &   26.38     &   27.38  &    41.03    &    39.47   \\ \hline \hline
SSB   &   0     &    0    &    0    &    0    &    0    &   0     &     0     &     0   &    0    &     0   &   0    &    0    &   0  &    0    &    0   \\
\end{tabular}
\label{tab:effectiveness_TGT}
\end{table*}
\begin{table*}
\vspace{-0.2cm}
\setlength{\abovecaptionskip}{0.05cm}
\setstretch{1}
\fontsize{6.2pt}{2.6mm}\selectfont
    \centering
    \caption{\small Effectiveness: relative error (\%) for different query shapes over all datasets and all queries (human-annotated ground truth)}
\begin{tabular}{c||c|c|c|c|c||c|c|c|c|c||c|c|c|c|c}
\multirow{2}{*}{\textbf{Method}} & \multicolumn{5}{c||}{\textbf{DBpedia}} & \multicolumn{5}{c||}{\textbf{Freebase}} & \multicolumn{5}{c}{\textbf{Yago2}} \\
 & Simple & Chain & Star & Cycle & Flower & Simple & Chain & Star & Cycle & Flower & Simple & Chain & Star & Cycle & Flower  \\ \hline \hline
\textbf{Ours}   &    \textbf{0.99 }      &   \textbf{0.84}     &   \textbf{1.10}     &    \textbf{1.19}      &   \textbf{0.80}    &    \textbf{0.96}    &  \textbf{1.38}    &  \textbf{1.02}    &  \textbf{0.98}     &     \textbf{1.60}     &    \textbf{0.57}    &    \textbf{1.30}    &   \textbf{0.86}       &   \textbf{0.92}      &   \textbf{1.04}      \\ \hline
EAQ    &    21.14       &    -    &    -    &    -    &    -    &   17.74     &    -      &     -   &   -    &     -     &  14.72    &    -    &    - &      -  &    -     \\ \hline
GraB  &     7.31       &   29.38     &    18.29    &    18.72    &    16.06    &   11.41     &     17.74     &     18.52   &    22.97    &    20.60    &   7.62    &       20.56 &   17.16  &   24.58     &   14.02    \\ \hline
QGA  &    19.01     &    32.90    &    40.52    &     25.58   &   32.45    &   34.65    &     29.65     &     34.68   &   21.53     &   22.27     &   18.94    &       23.81 &   32.70  &     43.50   &  34.35     \\ \hline
SGQ   &     9.97     &   12.93     &    17.16    &   9.08     &    11.64    &   5.30     &     4.33     &     14.16   &   12.54     &    17.28    &   7.01    &    13.19    &   9.35  &   15.31     &   9.63    \\ \hline
JENA &    17.62     &    30.79    &    44.31    &    22.77    &    21.22    &   27.66     &     29.65     &     35.39   &    22.58    &    23.02    &   12.40    &   26.19     &   27.54  &   42.23     &    41.69   \\ \hline
Virtuoso &  17.62     &    30.79    &    44.31    &    22.77    &    21.22    &   27.66     &     29.65     &     35.39   &    22.58    &    23.02    &   12.40    &   26.19     &   27.54  &   42.23     &    41.69   \\ \hline \hline
SSB   &   0.91     &    0.81    &    0.92    &    1.10    &    0.73    &   1.83     &     1.32     &     0.93   &    0.91    &     1.70   &   1.11    &    1.21    &   0.83  &    0.85    &    1.13   \\
\end{tabular}
\label{tab:effectiveness_HAGT}
\end{table*}
\begin{table*}
\vspace{-0.2cm}
\setlength{\abovecaptionskip}{0.05cm}
\setstretch{1}
\fontsize{6.2pt}{2.6mm}\selectfont
    \centering
    \caption{\small Efficiency: average response time (ms) for different query shapes over all datasets and all queries}
\begin{tabular}{c||c|c|c|c|c||c|c|c|c|c||c|c|c|c|c}
\multirow{2}{*}{\textbf{Method}} & \multicolumn{5}{c||}{\textbf{DBpedia}} & \multicolumn{5}{c||}{\textbf{Freebase}} & \multicolumn{5}{c}{\textbf{Yago2}} \\
 & Simple & Chain & Star & Cycle & Flower & Simple & Chain & Star & Cycle & Flower & Simple & Chain & Star & Cycle & Flower  \\ \hline \hline
\textbf{Ours}   &    \textbf{368.0 }      &   \textbf{590.63}     &   \textbf{735.1}     &    \textbf{841.9}      &   \textbf{1518.3}    &    \textbf{419.4}    &  \textbf{673.1}    &  \textbf{905.8}    &  \textbf{1037.7}     &     \textbf{1871.1}     &    \textbf{562.5}    &    \textbf{902.8}    &   \textbf{1154.4}       &   \textbf{1322.4}      &   \textbf{2384.6}      \\ \hline
EAQ    &    4525.3       &    -    &    -    &    -    &    -    &   2711.0     &    -      &     -   &   -    &     -     &  4267.4    &    -    &    - &      -  &    -     \\ \hline
GraB  &     922.6       &   1541.3     &    1918.1    &    1727.4    &   3114.9     &   607.2     &     1014.5     &     1471.5  &    1325.2    &    2389.6    &   1449.0    &    2420.9    &  3511.5  &   3162.4     &   5702.5    \\ \hline
QGA  &    1511.0     &   2120.7     &    2239.1    &    3023.3    &    5451.6    &   1088.5    &    1527.8      &     2173.3  &     2489.7   &    4489.4    &  2369.8    &     3326.1   &   2703.1  &    3096.6    &   5583.8    \\ \hline
SGQ   &     817.4     &    1033.2    &    1285.7    &    1157.9    &     2087.9   &   621.9     &      786.1    &     1174.5   &     1057.8   &    1907.4    &   1392.5    &   1760.1     &  2630.2  &    2368.5    &   4271.1    \\ \hline
JENA &    1197.9     &    1681.4    &    2546.2    &   2916.9     &    5259.8    &   1051.4     &    1475.7      &     1606.1   &    1839.9    &    3317.7    &   1505.7    &    2113.3    &   2702.2  &    3095.5    &   5581.9    \\ \hline
Virtuoso & 1212.2     &   1701.4     &    2550.9    &    2922.2    &    5269.4    &   1074.8     &    1508.5      &     1634.6   &    1872.5    &    3376.6    &   1545.9     &    2169.7    &   2894.4   &     3315.7   &   5979.1    \\ \hline \hline
SSB   &   6675.8      &    7547.9     &    9577.6    &    10972.3    &    20578.6    &   2264.0      &     4766.5     &     10003.1    &   11459.7     &      20663.5 &   5231.1     &    6627.6    &   9624.5   &    10471.2    &    19278.6   \\
\end{tabular}
\label{tab:efficiency}
\vspace{-0.5cm}
\end{table*}

\vspace{0.05cm}
\noindent\underline{\textbf{Comparing methods.}} We compared our approach with recent works on KG search: (1) EAQ \cite{Li2020} is state-of-the-art work that supports aggregate queries on KGs: It collects candidate entities via link prediction and computes aggregate results {\em only for simple queries} (\S \ref{previous_work}). (2) SGQ \cite{Wang2020} finds the top-$k$ semantically similar answers for a query graph in a KG. SGQ supports incremental query processing: When $k$ is increased, the additional answers can be retrieved incrementally, without finding all the top-$k$ answers from ground. To find all correct answers, we initialize $k$=50 and then increase $k$ in steps of 50 till all correct answers are included. Notice that in this process, some incorrect answers can also get included in the last step. (3) GraB \cite{Jin2015} is an index-free graph query method based on structural similarity. (4) QGA \cite{Han2017} is a keyword based KG search method. We also compared with one {\em RDF store and a graph DB, both supporting SPARQL queries}: (5) JENA \cite{Jena} is a well-known RDF store and (6) Neo4j \cite{Neo4j} is a famous graph DBMS, on which we deploy a sparql-plugin. We also compared with (7) SSB provided in \S \ref{exact}. Since (2)-(4) process factoid queries, we extended them by adding an additional aggregate operation after achieving the factoid query answers. Moreover, (7) serves as a baseline method, we demonstrate our accuracy vs. efficiency trade-offs by comparing with (7), thereby showing the usefulness of ``sampling-estimation''.

\vspace{0.05cm}
\noindent\underline{\textbf{Parameters.}} The default configuration is: error bound $e_b$ = $1\%$, confidence level $1-\alpha$ = $95\%$, repeat factor $r$ = $3$, desired sample ratio $\lambda$ = $0.3$, $n$ = $3$ for $n$-bounded subgraph, similarity threshold $\tau$ from Table \ref{tab:jaccard}, and TransE \cite{Bordes2013} for KG embedding.

\vspace{0.05cm}
\noindent\underline{\textbf{Remarks.}} {\bf (1)} In real applications, HA-GT may not be available for all queries. Alternately, a domain expert can set $\tau$ according to her experience and based on available human-annotations for a limited number of queries, and then use our solution to retrieve a good approximation to both $\tau$-GT (Table \ref{tab:effectiveness_TGT}) and HA-GT (Table \ref{tab:effectiveness_HAGT}) over all queries. The key is whether we can find an appropriate $\tau$ that makes the human-annotated correct answers and $\tau$-relevant correct answers highly consistent. {\bf (2)} Table \ref{tab:jaccard} shows the average Jaccard similarity (AJS) between the $\tau$-relevant (with different $\tau$) and human-annotated correct answers and the variance (Var) of Jaccard similarity, considering 35\% of all queries over three datasets (with TransE KG embedding). Even though we have HA-GT for {\em all} queries with additional
crowdsourcing, in Table \ref{tab:jaccard} we use HA-GT from 35\% of all queries and find the appropriate $\tau$ for each dataset. This is to demonstrate that the optimal $\tau$ obtained based on available human-annotations for {\em a limited number of queries} on a dataset, is still sufficient to retrieve a good approximation to both $\tau$-GT and HA-GT for many other queries on that dataset. Intuitively, if we have a high-quality KG embedding model, then we can distinguish the implicit semantics of predicates well; hence, we can also accurately represent the semantics of different paths. So, it is very likely that the ground truths HA-GT and $\tau$-GT would be similar for an appropriate value of $\tau$.
{\bf (3)} The optimal $\tau$ for different datasets are different; though roughly in the range of 0.8-0.85, we can get a relatively higher AJS and smaller Var. For example, in DBpedia, AJS is the highest (0.95) and Var is the smallest (0.011) when $\tau=0.85$, implying that the difference of each query's $\tau$-relevant (for $\tau=0.85$) correct answers to its human annotated correct answers is small. In this case, our solution can get a good approximation to HA-GT (Table \ref{tab:effectiveness_HAGT}), because HA-GT and $\tau$-GT are similar when $\tau=0.85$, and our solution can achieve a good approximation to $\tau$-GT (Table \ref{tab:effectiveness_TGT}).

\subsection{Effectiveness and Efficiency Evaluation}
\label{effective}
\noindent\underline{\textbf{Effectiveness.}} Table \ref{tab:effectiveness_TGT} and \ref{tab:effectiveness_HAGT} show the relative error based on $\tau$-GT (with the optimized configuration of $\tau$ from Table \ref{tab:jaccard}) and HA-GT, respectively, for {\em all queries} having different graph shapes. Our solution achieves a good approximation to $\tau$-GT (Table \ref{tab:effectiveness_TGT}). In Table \ref{tab:effectiveness_TGT}, {\em for all datasets, our solution has two orders of magnitude less relative error on average than other methods}. The reasons are: {\bf (1)} We collected the random sample $S_\mathcal{A}$ by considering the semantic similarity via ``edge-to-path" mapping in the KG, so that we could find more correct answers than other graph query methods (EAQ, GraB, QGA) that do not consider semantic similarity. {\bf (2)} {\em Since we provided a specific SPARQL expression as input to both JENA and Neo4j, they only found those correct answers matching {\em exactly} with the graph schema of the input SPARQL query, and other correct answers having different schemas were ignored}. {\bf (3)} Our approach continues to refine the approximate result by increasing $S_\mathcal{A}$ until the relative error is bounded by the user-specified input error bound $e_b$. We show a case study for Q1, Q2, and Q6 in Table \ref{tab:casestudy_effectiveness}, where the approximate results are refined iteratively until all relative errors are bounded by $e_b=1\%$. {\bf (4)} For the top-$k$ based incremental method SGQ, its relative error $>0$. This is because we increased $k$ in steps of 50 till all correct answers were included, in this process, some incorrect answers were included in the last step.

\begin{table}
\setstretch{1}
\setlength{\abovecaptionskip}{0.1cm}
\fontsize{6.2pt}{2.6mm}\selectfont
\begin{minipage}[!t]{.56\columnwidth}
  \caption{\footnotesize Case study: relative error (\%) \\refinement ($\tau$-GT for Q1, Q2, and Q6 are 596, \$44,072, and \$7.56B)}
  \label{tab:casestudy_effectiveness}
  \centering
\begin{tabular}{c||c|c|c|c}
\multirow{2}{*}{\textbf{QID}}  & \multicolumn{4}{c}{\textbf{Approximate result}} \\
\cline{2-5}
 & round   &      $\hat{V}$       &  \textbf{MoE} $\varepsilon$         & \tabincell{c}{error \%}  \\ \hline
\multirow{2}{*}{Q1}          & 1       & 578.55      & 8.51       & 2.93        \\ 
                                                                                & 2       & \textbf{599.97}      & \textbf{5.99}       &  \textbf{0.67}        \\ \hline
\multirow{2}{*}{Q2}   & 1       & 46,409   & 1,265   & 5.30        \\ 
                                                                                & 2       & \textbf{44,504}   & \textbf{435}     & \textbf{0.98}        \\ \hline
\multirow{3}{*}{Q6}  & 1       & 7.07        & 0.59       & 6.48        \\ 
                                                                    & 2      & 7.89       &  0.13   &     4.37  \\ 
                                                                   & 3       & \textbf{7.53}        & \textbf{0.071}       & \textbf{0.40}        \\ 
\end{tabular}
  \end{minipage}
  \hspace{0.08cm}
\begin{minipage}[!t]{.4\columnwidth}
  \caption{\footnotesize Efficiency for various operators ({\em DBpedia})}
  \label{tab:filter_efficiency}
  \centering
\begin{tabular}{c||c|c|c}
\multirow{2}{*}{\textbf{Method}}  & \multicolumn{3}{c}{\textbf{Efficiency (sec)}} \\
\cline{2-4}
& Filter & GROUP       & MAX/     \\
 &        & -BY     & MIN \\ \hline 
\textbf{Ours}           & \textbf{0.43}       & \textbf{31.67}      &  \textbf{0.47}       \\ \hline
EAQ			&  -  & -   & 2.68 \\ \hline
GraB		& 1.10    &    -    &   0.80   \\ \hline
QGA			&  1.41     &     -   &   1.28    \\ \hline
SGQ			& 0.73    &    -    &   0.64     \\ \hline
JENA		& 0.70     &   95.76     &  1.07    \\ \hline
Virtuoso	& 0.72    &   94.67     &   0.97   \\ \hline
SSB			& 0.78  &  54.75  &  8.17   \\
\end{tabular}
  \end{minipage}
\end{table}

Table \ref{tab:effectiveness_HAGT} shows the effectiveness results w.r.t. HA-GT for {\em all queries}. By setting an appropriate $\tau$, our solution can get a good approximation to HA-GT, because HA-GT and $\tau$-GT are similar for this $\tau$. However, more results have slightly larger relative error than the predefined error bound ($e_b=1\%$). This is reasonable, since we provide an accuracy guarantee for $\tau$-GT, and not directly for HA-GT; though HA-GT and $\tau$-GT are similar for a specific $\tau$, they may not be exactly same.
Moreover, if we do not select $\tau$ appropriately, the relative error w.r.t. HA-GT could be affected. We show the effect of $\tau$ w.r.t. HA-GT in \S \ref{sensitivity} (Figure \ref{fig:threshold} (right)).

Table \ref{tab:filter} shows the effectiveness results w.r.t. HA-GT and $\tau$-GT for queries with filters, GROUP-BY, 
and MAX/MIN. 
For filters and GROUP-BY, we can achieve \textit{two orders of magnitude less error on average than others}, because we offer a good accuracy guarantee with confidence interval. For MAX/MIN, although we cannot provide accuracy guarantees, we can support them by simply returning the MAX/MIN answers of the collected sample. As more sample is collected, the result gets closer to the exact one. We configured a fix sample size (5\% of the candidate answers) and found that {\em the exact result can be included in the sample after 8 rounds on average}. We show the result after 4 rounds, which is already better than others.

\begin{table}
\setlength{\abovecaptionskip}{0.05cm}
\setstretch{1}
\fontsize{6.2pt}{2.6mm}\selectfont
    \centering
    \caption{\small Effectiveness for various operators (\footnotesize \textit{DBpedia}, HA-GT)}
\begin{tabular}{c||c|c|c||c|c|c}
\multirow{2}{*}{\textbf{Method}} & \multicolumn{3}{c||}{\textbf{Relative error (\%) w.r.t. $\tau$-GT}} & \multicolumn{3}{c}{\textbf{Relative error (\%) w.r.t. HA-GT}} \\
\cline{2-7}
 & Filter & GROUP-BY & MAX/MIN & Filter & GROUP-BY & MAX/MIN \\ \hline \hline
\textbf{Ours}   &    \textbf{0.58 }      &   \textbf{0.75}     &   \textbf{5.68}     &    \textbf{0.71}      &   \textbf{1.13}    &    \textbf{6.33}    \\ \hline
EAQ    &    -       &    -    &    10.29    &    -    &    -  &  10.92   \\ \hline
GraB  &     21.49       &    -    &    10.02    &    20.94    &    -    &   10.65   \\ \hline
QGA  &    45.55     &    -    &    11.23    &   46.95     &     -   &   11.85    \\ \hline
SGQ   &     18.42     &     -   &    6.14    &    17.71    &    -    &   6.79     \\ \hline
JENA &    46.18     &   16.75     &    12.59    &   48.98     &   16.30     &  13.21    \\ \hline
Virtuoso &  46.18     &   16.75     &    12.59    &   48.98     &   16.30     &  13.21   \\ \hline
SSB &  0     &   0     &    0    &   1.29     &   1.06     &  1.14   \\
\end{tabular}
\label{tab:filter}
\vspace{-0.4cm}
\end{table}

\begin{table}
\vspace{0.2cm}
\setstretch{1}
\setlength{\abovecaptionskip}{0.1cm}
\fontsize{6.3pt}{2.8mm}\selectfont
\begin{minipage}[!t]{.47\columnwidth}
  \caption{\footnotesize Detailed efficiency results (ms) over (\textit{DBpedia}, simple)}
  \label{tab:detailed_efficiency}
  \centering
\begin{tabular} {c||c|c|c}
   \textbf{Operator} & \textbf{S1} & \textbf{S2} & \textbf{S3}  \\
   \hline 
   COUNT & 246.0 & 19.6 & 6.2  \\ 
   AVG & 248.2 & 104.1 & 42.4  \\ 
   SUM & 277.1 & 108.9 & 51.5
  \end{tabular}
  \end{minipage}
  \hspace{0.08cm}
\begin{minipage}[!t]{.52\columnwidth}
  \caption{\footnotesize Effect of KG embedding models (\textit{DBpedia}, simple, HA-GT)}
  \label{tab:embedding}
  \centering
\begin{tabular} {c||c|c|c}
   \textbf{Model} & \textbf{Embed}    & \textbf{Mem} & \textbf{Relative} \\
                  & \textbf{time (h)} & \textbf{(GB)}& \textbf{error(\%)} \\
   \hline 
   TransE & {\bf 6.63} & {\bf 8.8} & {\bf 0.99}   \\ 
   TransD & 10.06 & 9.73 & {\bf 0.83}  \\ 
   TransH & 7.82 & 9.35 & 1.07 \\
   RESCAL & $\approx$1day & 50 & 5.46 \\
   SE & $\approx$1day & 55 & 3.38
  \end{tabular}
  \end{minipage}
\end{table}

\begin{figure*}
\vspace{-0.9cm}
\setlength{\abovecaptionskip}{-0.1cm}
\centering
\subfigcapskip=-0.18cm
\subfigure[\small Effect of S1 on effectiveness, efficiency]{
\captionsetup{skip=0pt}
\includegraphics[scale=0.36]{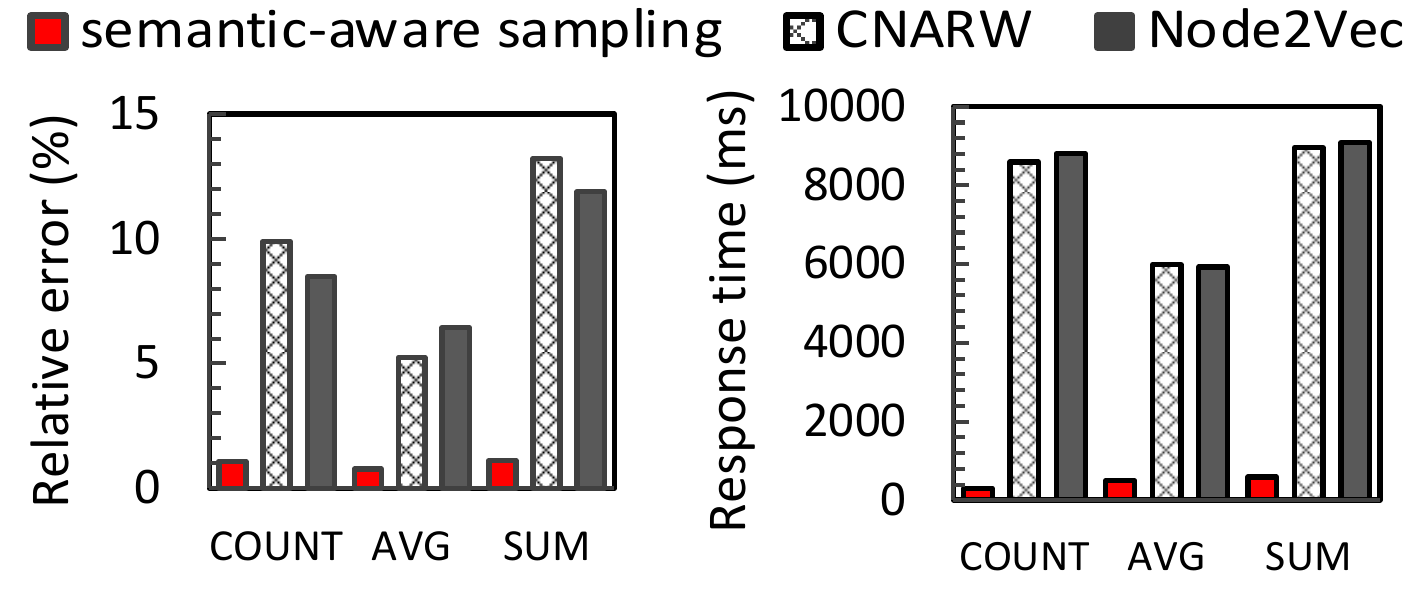}
\label{fig:effective_s1}
}
\subfigure[\small Effect of S2 on effectiveness, efficiency]{
\captionsetup{skip=0pt}
\vspace{0mm}
\includegraphics[scale=0.36]{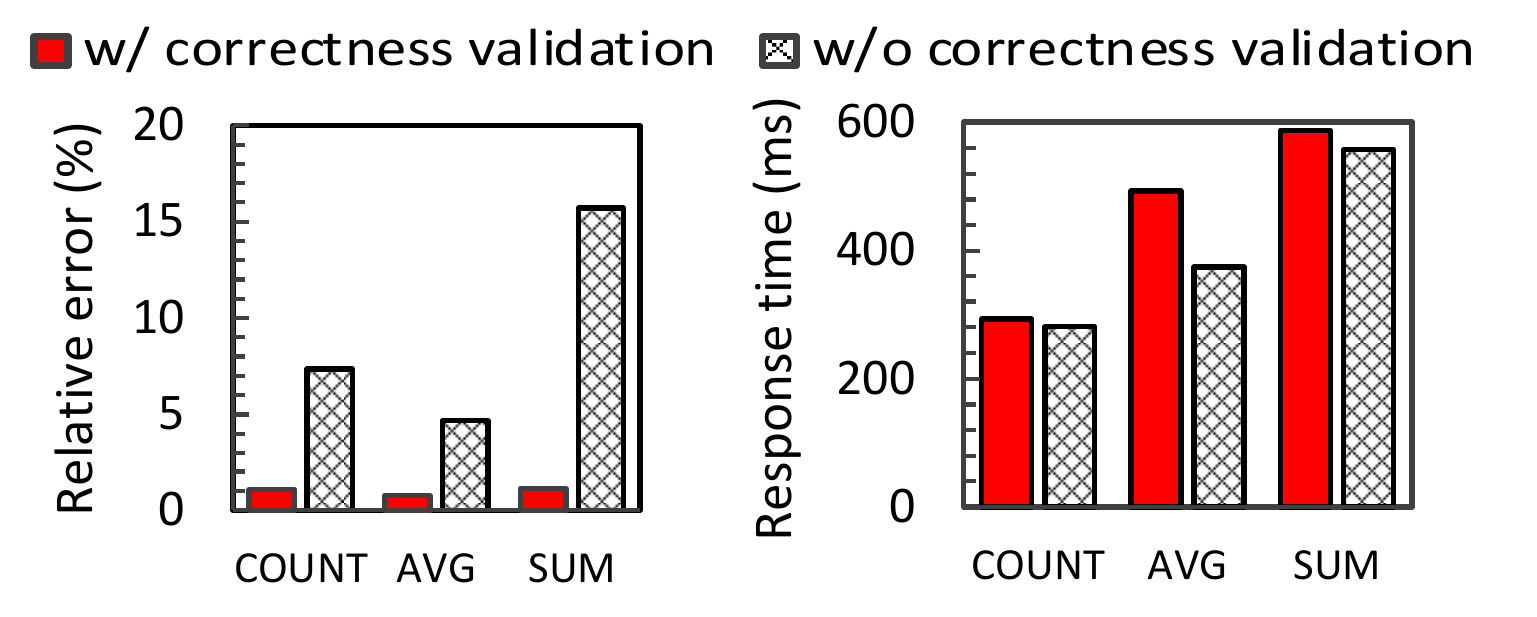}
\label{fig:effective_s2}
}
\subfigure[\small Effect of S3 on effectiveness, efficiency]{
\captionsetup{skip=0pt}
\vspace{0mm}
\includegraphics[scale=0.36]{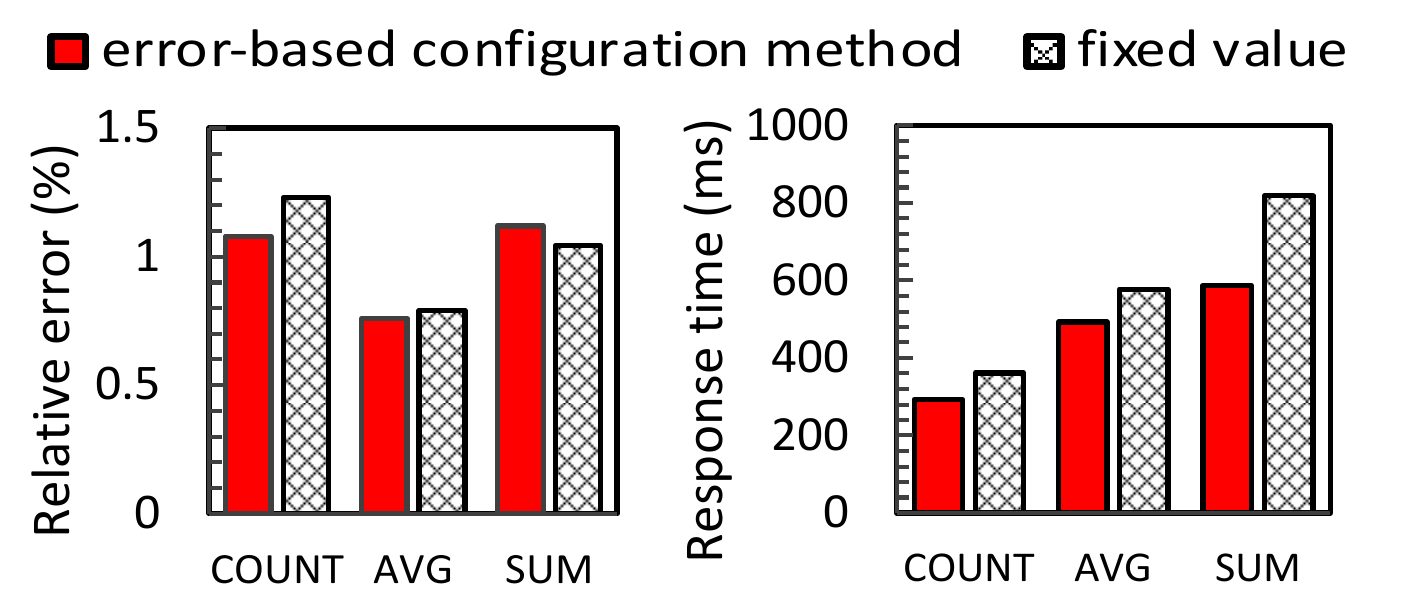}
\label{fig:effective_s3}
}
\caption{\small Effect of each step (S1-S3) on effectiveness and efficiency ({\em DBpedia}, simple, HA-GT)}\vspace{-0.4mm}
\label{fig:effect_step}
\vspace{-0.15cm}
\end{figure*}

\begin{figure*}
\setlength{\abovecaptionskip}{0.1cm}
\vspace{-0.3cm}
\centering
\subfigcapskip=-0.32cm
\subfigure[\small Interactive performance]{
\captionsetup{skip=0.1pt}
\includegraphics[scale=0.35]{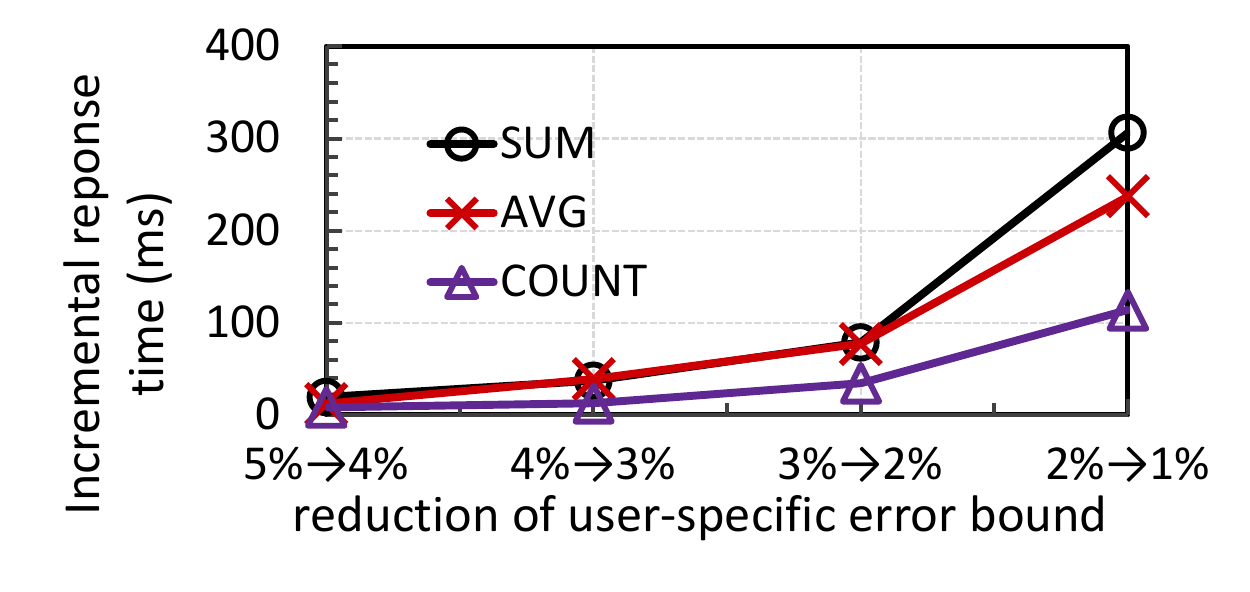}
\label{fig:errorbound}
\vspace{-0.3cm}
}
\subfigure[\small Effect of confidence level $1-\alpha$]{
\captionsetup{skip=0.1pt}
\includegraphics[scale=0.35]{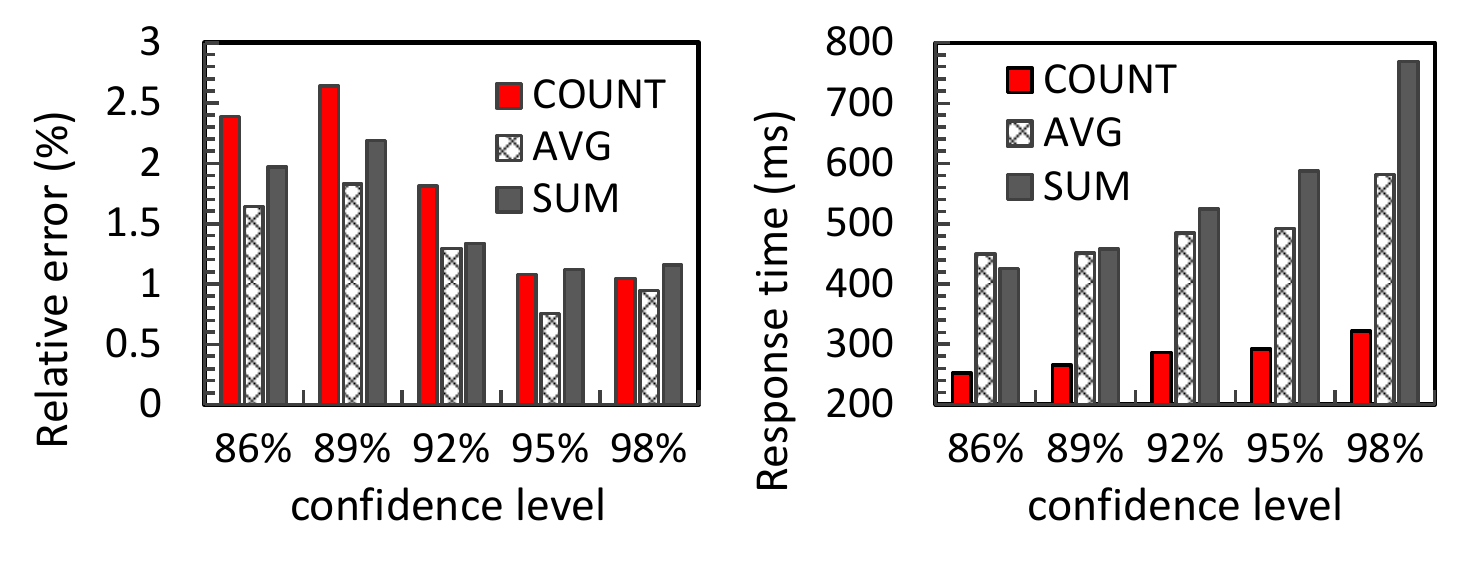}
\label{fig:confidencelevel}
}\vspace{-0.4cm}
\subfigure[\small Effect of repeat factor $r$]{
\captionsetup{skip=0.1pt}
\includegraphics[scale=0.35]{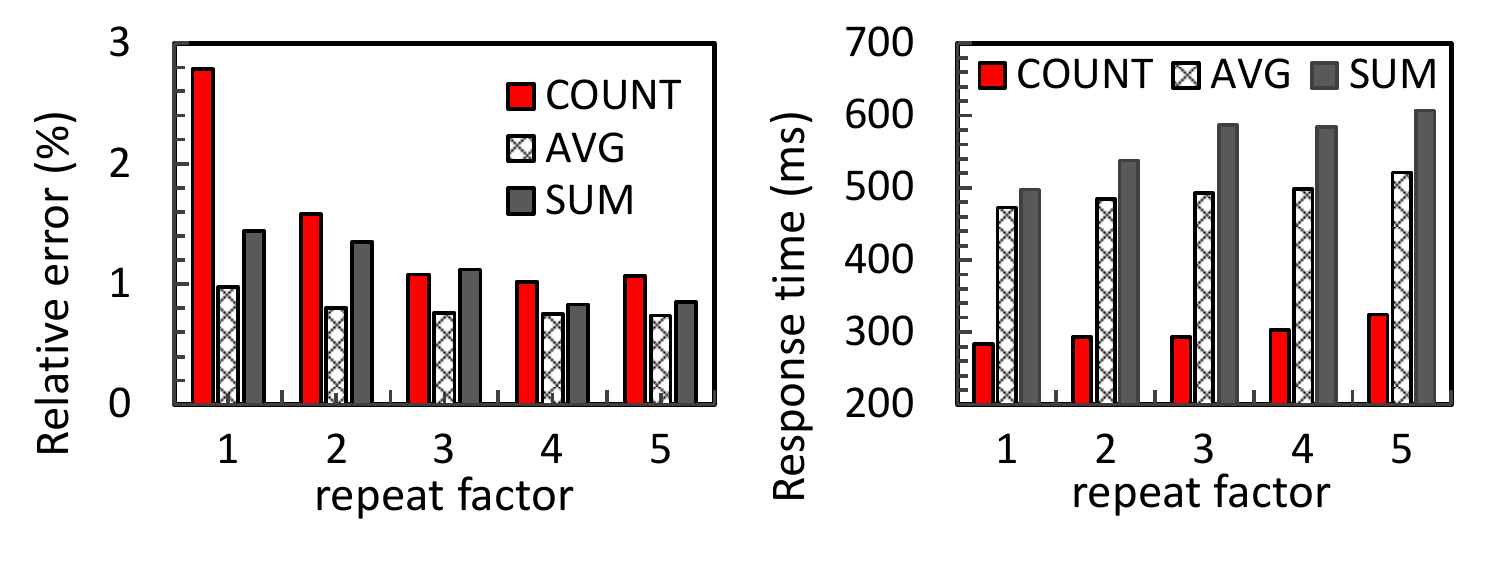}
\label{fig:repeatfactor}
}
\subfigure[\small Effect of desired sample ratio $\lambda$]{
\captionsetup{skip=0.1pt}
\includegraphics[scale=0.35]{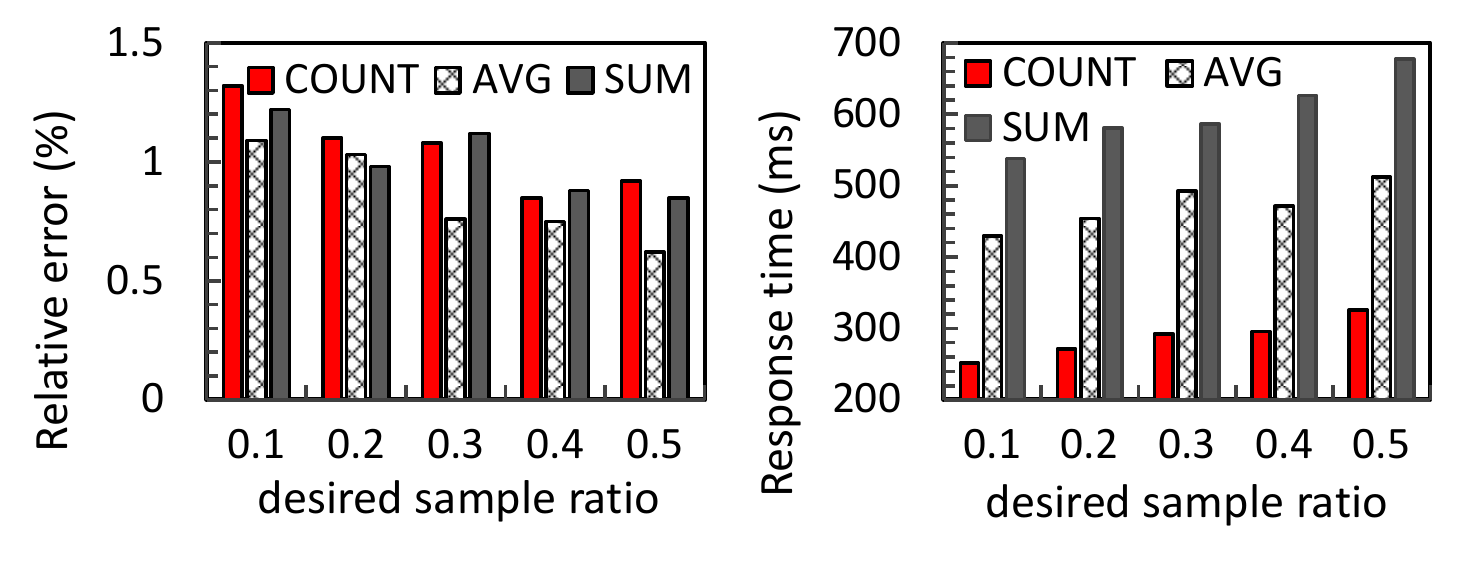}
\label{fig:desiredsampleratio}
}\vspace{-0.1cm}
\subfigure[\small Effect of $n$-bounded subgraph]{
\captionsetup{skip=0.1pt}
\includegraphics[scale=0.35]{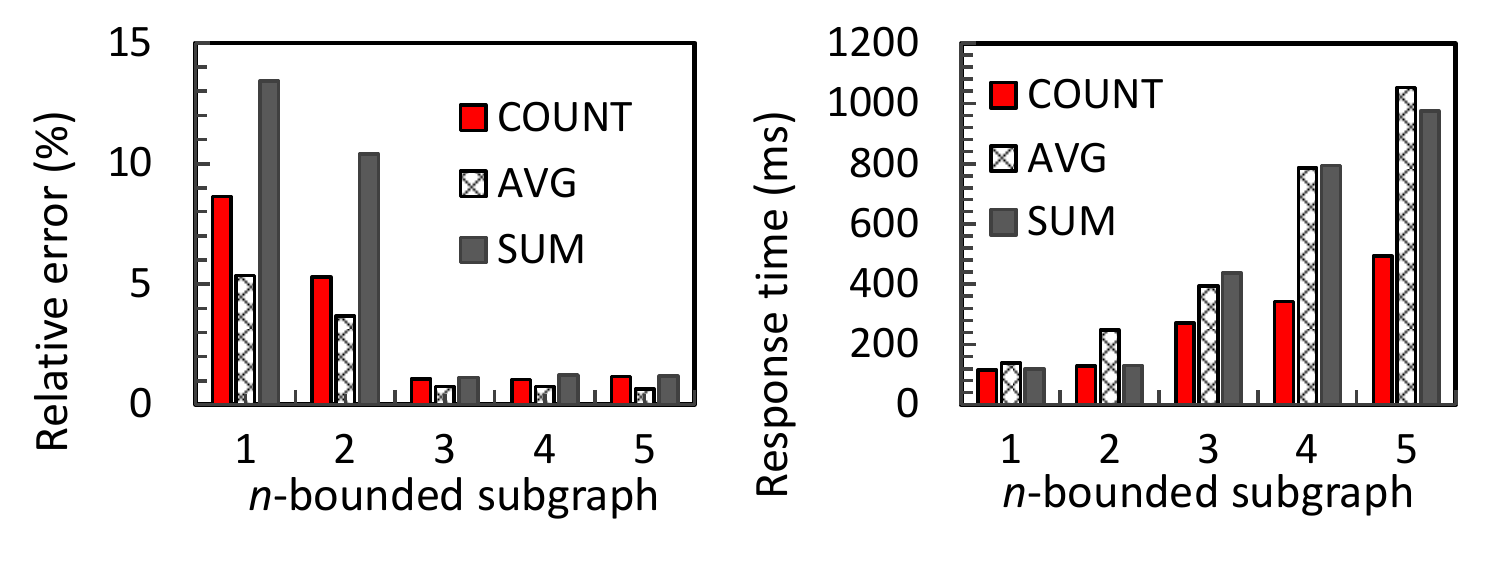}
\label{fig:nboundedgraph}
}
\subfigure[\small Effect of semantic similarity threshold $\tau$]{
\captionsetup{skip=0pt}
\includegraphics[scale=0.35]{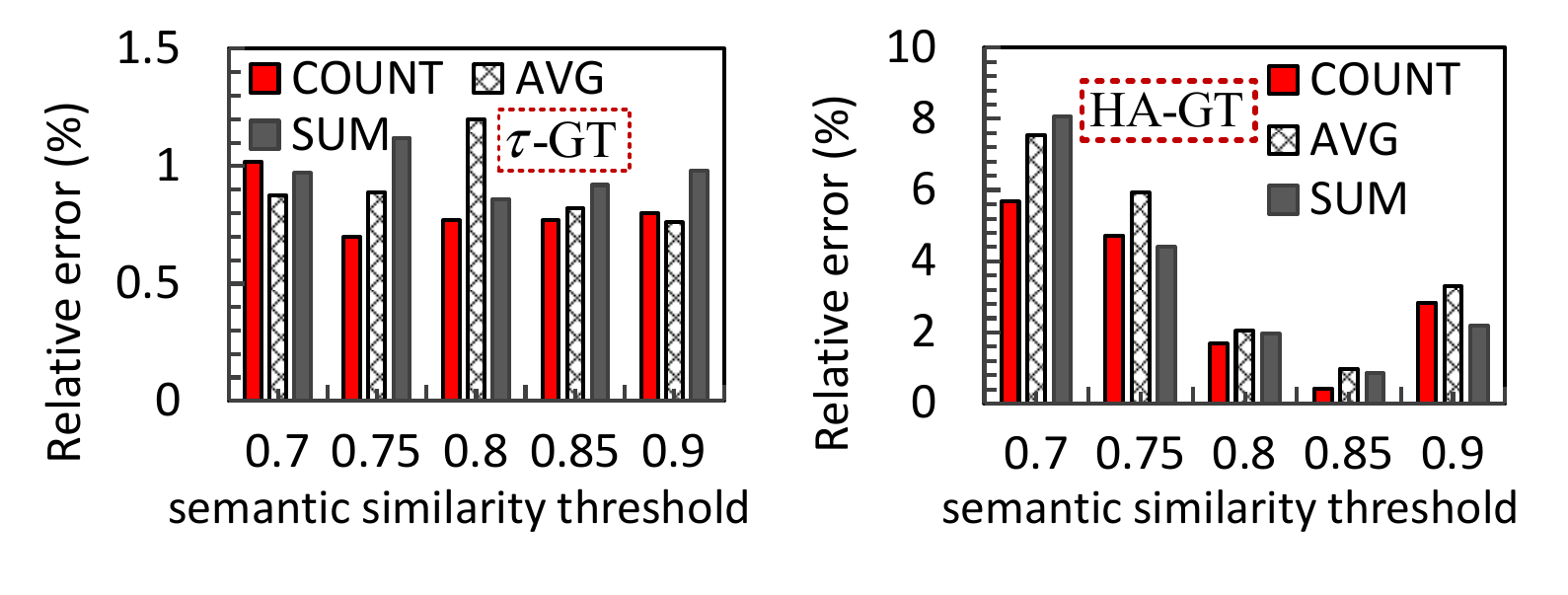}
\label{fig:threshold}
}
\caption{\small (a) Interactive performance; (b)-(f) Effect of different parameters on effectiveness and efficiency ({\em DBpedia}, simple, HA-GT)}
\label{fig:parameter}
\vspace{-0.5cm}
\end{figure*}

\vspace{0.05cm}
\noindent\underline{\textbf{Efficiency.}} The runtime of our method is independent of which ground truth is used. Specifically, when the query terminates, we use the approximate result to compute the relative error w.r.t. HA-GT and $\tau$-GT to obtain different effectiveness results, while the query time is the same. Hence, we report the efficiency results in Tables \ref{tab:efficiency} and \ref{tab:filter_efficiency}, and find that {\em our method requires up to an order of magnitude less response time than others}, because we apply the ``sampling-estimation" model that does not rely on the factoid query. Hence, we do not need to wait a long time for the graph query results, thereby improving the efficiency. The query time increases as the graph shape gets more complex. This is because we use the ``decomposition-assembly" framework to answer complex queries -- with more sub-queries, more time is required for sampling and correctness validation. We show the detailed efficiency results of our three steps on DBpedia in Table \ref{tab:detailed_efficiency}: \textit{semantic-aware sampling} (S1), \textit{approximate estimation} (S2), and \textit{accuracy guarantee} (S3). S1 is the most time-consuming step and S3 is the fastest step, because we must wait for the more time-consuming random walk to converge before collecting the sample. In S2, we validate the correctness of the collected sample, this costs more time than that for S3.
%

\subsection{Effect of Each Step on the Performance}
\label{scale}

\vspace{0.05cm}
\noindent\underline{\textbf{Semantic-aware sampling (S1).}} We implemented other two versions of S1: both topology-aware samplings (CNARW \cite{Li2019} and Node2Vec \cite{Grover2016}). Figure \ref{fig:effective_s1} shows 
that the relative errors are reduced by at least 8X, 6X, and 10X for ${\sf COUNT}$, ${\sf AVG}$, and ${\sf SUM}$, respectively, by using our semantic-aware sampling. 
Our method is also more efficient, because our semantic-aware sampling converges faster than CNARW and Node2Vec.

\vspace{0.05cm}
\noindent\underline{\textbf{Approximate estimation (S2).}} In S2, we deployed a correctness validation method to identify the correctness of answers in a sample.
We show the effectiveness result w/ or w/o correctness validation in Figure \ref{fig:effective_s2} (left). The relative errors are reduced by 7X, 6X, and 14X for ${\sf COUNT}$, ${\sf AVG}$, and ${\sf SUM}$, through correctness validation. Figure \ref{fig:effective_s2} (right) shows that the correctness validation increases the response time, but it is acceptable considering the improvement in accuracy.

\vspace{0.05cm}
\noindent\underline{\textbf{Accuracy guarantee (S3).}} In S3, We applied an error-based method to automatically configure $|\Delta S_\mathcal{A}|$ to refine the result iteratively. Different to this, many approximate query processing methods on relational dataset increase the sample size with a fixed value, e.g., 50. 
Figure \ref{fig:effective_s3} shows that the relative errors are similar for both, but ours offers better efficiency (improved by 1.3X on average), as we can automatically configure $|\Delta S_\mathcal{A}|$, avoiding oversampling or undersampling.
%

\subsection{Interactive Performance and Parameter Sensitivity}
\label{sensitivity}

We studied our interactive approach's runtime by varying the user-specific error bound $e_b$ during query processing. We initialize $e_b=5\%$. Instead of terminating the query when $\varepsilon\leq \hat{V}\cdot e_b/(1+e_b)$, we keep decreasing $e_b$ by $1\%$ to continue processing and report the incremental response time.
Figure \ref{fig:errorbound} shows
that we require less than 100 ms of additional time to meet a new $e_b$ till $e_b\geq 2\%$, which is efficient. While for a more stringent accuracy requirement such as $e_b=1\%$, we only need 100$\sim$300 ms to re-satisfy the termination condition.

\vspace{0.05cm}
\noindent\underline{\textbf{KG embedding.}} 
We measure query's accuracy (based on HA-GT) due to different KG embedding models (Table \ref{tab:embedding}):
a series of translation-based models (TransE \cite{Bordes2013}, TransD \cite{Ji2015}, TransH \cite{Wang2014}),
tensor factorization-based model RESCAL \cite{NickelTK11}, and relation-specific projection-based model SE \cite{BordesWCB11}.
Different models can preserve different types of relation properties, e.g., symmetry, antisymmetry, inversion,
composition, complex mapping properties, etc. \cite{SunDNT19}. Unlike RESCAL and SE,
translation-based models preserve many important properties
such as antisymmetry, inversion, and composition, which are critical for the datasets and queries that we tested,
so these models can well-represent the semantics of predicates and paths, and we find a
good $\tau$ which makes $\tau$-GT obtained from these models similar to HA-GT.
Therefore, translation-based models perform better than others.
We find that different translation-based models (TransE, TransD, TransH)
have small differences in accuracy. TransE is also more efficient in embedding time and memory (Table \ref{tab:embedding}).
These results demonstrate that a high-quality KG embedding model is important to our solution when we use HA-GT.

\vspace{0.05cm}
\noindent\underline{\textbf{Confidence level $1-\alpha$.}} 
Relative error reduces as  $1-\alpha$ increases (Fig. \ref{fig:confidencelevel}): The higher $1-\alpha$ is, the tighter MoE $\varepsilon$ is (Eq. \ref{eq:moe}), i.e., the ground truth $V$ is covered by a narrow CI with a higher confidence level. Response time increases with $1-\alpha$, as we need more time and samples to update $\varepsilon$. 


\vspace{0.05cm}
\noindent\underline{\textbf{Repeat factor $r$.}} As Figure \ref{fig:repeatfactor} shows, 
a larger $r$ improves the precision of correctness validation by reducing false negatives. This improvement gets stable when $r$ $>$ $3$. 

\vspace{0.05cm}
\noindent\underline{\textbf{Desired sample ratio $\lambda$.}} The larger $\lambda$ is, the larger sample $S_\mathcal{A}$ is collected,
and the relative error decreases (Figure \ref{fig:desiredsampleratio}). 
However, conducting the BLB over a larger
$S_\mathcal{A}$ is more time-consuming.
We achieve a balance with $\lambda=30\%$.

\vspace{0.05cm}
\noindent\underline{\textbf{$n$-bounded subgraph search.}} 
Figure \ref{fig:nboundedgraph} shows that the relative error decreases as $n$ increases.
Since most correct answers are included in the $3$-bounded subgraph, the reduction of relative error gets stable when $n\geq 3$.
Moreover, the query time increases as $n$ increases, because we need more time to do random walk on a larger subgraph until it converges.

\vspace{0.05cm}
\noindent\underline{\textbf{Semantic similarity threshold $\tau$.}} With semantic similarity-based ground truth, we achieve
good approximate results, i.e., relative error$<$1.5\%
(Fig. \ref{fig:threshold}(left)), since we sample answers having higher semantic similarity (\S \ref{sampling}), validate answers' correctness (\S \ref{similarity}),
and iteratively estimate the approximate result based on new samples until we obtain an accurate enough result (\S \ref{accuracy}).
For human-annotated ground truth (Fig. \ref{fig:threshold}(right)), we should carefully select $\tau$, we achieve the smallest relative error when $\tau$ $=$ $0.85$
for {\em DBpedia}, because semantic similarity-based and human-annotated ground truths are quite similar when $\tau$ $=$ $0.85$ with {\em DBpedia} (Table \ref{tab:jaccard}). 
\section{Conclusions}
\label{conclusion}
We proposed a ``sampling-estimation'' model to answer aggregate queries on KGs effectively and efficiently. We first presented a semantic-aware sampling to collect a high-quality sample from a KG. Then, we proposed two unbiased estimators for ${\sf COUNT}$ and ${\sf SUM}$,
and a consistent estimator for ${\sf AVG}$ to compute the approximate result. An effective accuracy guarantee was provided through a tight confidence interval, that is, the relative error of approximate result is bounded by a user-input error bound. We extended our solution for iterative improvement of accuracy, complex queries with filters, GROUP-BY, and different shapes. The experimental results on real datasets confirm the effectiveness and efficiency of our approach. In future, we would
derive accuracy guarantees for extreme functions (e.g., ${\sf MAX}$, ${\sf MIN}$) in our framework.


\bibliographystyle{IEEEtran}
\bibliography{RDF}

\end{document}